\documentclass[aps,pra,twocolumn,superscriptaddress,floatfix]{revtex4-2}

\usepackage{amssymb,amsmath}
\usepackage{mathtools}
\usepackage{graphicx}
\usepackage{dcolumn}
\usepackage{bm}
\usepackage[dvipsnames]{xcolor}
\usepackage{hyperref}
\hypersetup{colorlinks=false,
            pdfborder={0 0 0}}
\usepackage{environ}
\usepackage{braket}
\usepackage[normalem]{ulem}

\newcommand{\Ucal}{2.57(1)}
\newcommand{\Uiii}{2.79(2)}
\newcommand{\Uiv}{3.69(2)}
\newcommand{\Uv}{4.67(3)}
\newcommand{\Uvi}{5.72(3)}
\newcommand{\Uviii}{7.00(4)}
\newcommand{\Uxi}{10.29(6)}

\newcommand{\Twidhot}{0.35(1)}
\newcommand{\Twidhotn}{0.35}
\newcommand{\Twidcold}{0.15(1)}

\newcommand{\Uqmccal}{2.83(4)}
\newcommand{\Uqmciii}{3.08(4)}
\newcommand{\Uqmciv}{4.08(6)}
\newcommand{\Uqmcv}{5.16(7)}
\newcommand{\Uqmcvi}{6.32(9)}
\newcommand{\Uqmcviii}{7.7(1)}
\newcommand{\Uqmcxi}{11.4(2)}

\makeatletter
\newwrite\remember@figures
\AtBeginDocument{\InputIfFileExists{\jobname.dft}{}{}\immediate\openout\remember@figures=\jobname.dft}
\AtEndDocument{\immediate\closeout\remember@figures}
\NewEnviron{dfigure}[1]{%
    \immediate\write\remember@figures{%
        \noexpand\rememberfigure{#1}{\unexpanded\expandafter{\BODY}}%
    }%
}
\NewEnviron{dfigure*}[1]{%
    \immediate\write\remember@figures{%
        \noexpand\rememberfiguretc{#1}{\unexpanded\expandafter{\BODY}}%
    }%
}
\newcommand{\placefigure}[2][tp]{%
    \csname remembered@figure@#2\endcsname{#1}%
}
\newcommand{\rememberfigure}[2]{%
    \global\@namedef{remembered@figure@#1}##1{%
        \begin{figure}[##1]#2\end{figure}%
    }%
}
\newcommand{\rememberfiguretc}[2]{%
    \global\@namedef{remembered@figure@#1}##1{%
        \begin{figure*}[##1]#2\end{figure*}%
    }%
}
\makeatother

\begin{document}

\title{Pseudogap in a Fermi-Hubbard quantum simulator}

\newcommand{\harvard}{Department of Physics, Harvard University, Cambridge, MA, USA}
\newcommand{\harvardqse}{Quantum Science and Engineering, Harvard University, Cambridge, MA, USA}
\newcommand{\jila}
{JILA, National Institute of Standards and Technology and the University of Colorado, Boulder, Colorado, USA}
\author{Lev Haldar Kendrick}
\author{Anant Kale}
\affiliation{\harvard}
\author{Youqi Gang}
\author{Alexander Dennisovich Deters}
\affiliation{\harvard}
\affiliation{\harvardqse}
\author{Martin Lebrat}
\affiliation{\harvard}
\affiliation{\jila}
\author{Aaron W. Young}
\author{Markus Greiner}
\affiliation{\harvard}

\begin{abstract}
Understanding doped Mott insulators is a fundamental goal in condensed matter physics, with relevance to cuprate superconductors and other quantum materials~\cite{anderson_1987,lee_wen_nagaosa_review,imada_metal_insulator_review}.
The doped Hubbard model minimally describes such systems, and has explicated some of their complex behavior~\cite{qin_computational_review,arovas_analytic_review}.
However, many open questions remain concerning the anomalous metallic states which emerge at low temperatures and intermediate doping and which, in cuprates, give rise to high-temperature superconductivity upon cooling~\cite{lee_wen_nagaosa_review,norman_pines_kallin_pseudogap,proust_taillefer_review}.
Here we observe a crossover between a normal metal and a pseudogapped metal in the Hubbard model by performing thermodynamic and spectroscopic measurements in a cold atom quantum simulator, leveraging a recent several-fold reduction in experimentally achievable temperatures~\cite{cooling_paper}.
Measurements of the compressibility show a maximum versus doping that develops upon cooling, signaling an inflection point in the equation of state.
We track this maximum versus interaction strength, revealing a line of thermodynamic anomalies in the phase diagram that separates an underdoped from an overdoped metal at large interactions.
Lattice modulation spectroscopy shows a loss of spectral weight at low energies in the underdoped regime which is non-uniform in the Brillouin zone, indicating the formation of a pseudogap.
We use this signal to establish a pseudogap phase diagram as a function of interactions and doping.
Our results experimentally demonstrate the existence of a pseudogapped metal in the Hubbard model, partially characterize the pseudogap regime, and suggest a link between the pseudogap and charge order which can be probed in future work.
Furthermore, this work demonstrates the utility of quantum simulation in addressing frontier problems in correlated electron physics.

\end{abstract}

\maketitle

\section*{Introduction}

The presence of a pseudogap has long challenged theories of high-$T_c$ superconductivity in cuprates.
The term `pseudogap' refers to a peculiar phenomenology observed in thermodynamic, spectroscopic, and transport data at small to intermediate dopings at temperatures above the superconducting $T_c$: though the system is metallic, many physical properties suggest a loss of density of states
~\cite{lee_wen_nagaosa_review,norman_pines_kallin_pseudogap,basov_electrodynamics,proust_taillefer_review,arpes_on_cuprates_review}.
This leads to qualitative departures from Fermi liquid theory~\cite{pines_nozieres}, preventing the application of conventional theories of superconductivity~\cite{tinkham_superconductivity}.
Proposed theoretical explanations for the pseudogap include singlet formation, highly renormalized Fermi liquids, unusual forms of symmetry breaking, and various types of topological order~\cite{monthoux_pines_1993,lee_wen_nagaosa_review,hidden_order_laughlin,wen_quantum_orders,varma_olc_2006,subir_ancilla,kyung_pseudogap_short_range}.
Still, the puzzle of the pseudogap metal remains a topic of intense theoretical and experimental study.

As the simplest theoretical model of a doped Mott insulator, the Hubbard model has become central to the study of correlated electrons, and hence of the pseudogap.
At strong interaction strengths, the model is known to host a Mott insulator at zero doping and a Fermi liquid at large doping.
At intermediate doping, its ground-state and low-temperature physics has been the focus of extensive computational studies \cite{qin_computational_review,arovas_analytic_review,leblanc_benchmark}.
These studies have established several ground state ordering tendencies, including stripe order and $d$-wave superconductivity~\cite{qin_computational_review}, as well as the presence of some pseudogap phenomena at low temperatures ~\cite{qin_computational_review,simkovic_pseudogap,wietek_metts_pseudogap,vilk_tremblay_1997,diagrammatic_dmft_review,schafer_footprints}.
At weak interaction strengths, it is known that long-ranged spin fluctuations can produce a pseudogap \cite{vilk_tremblay_1997,schafer_footprints}.
However, the  mechanism for the pseudogap at strong interactions, where the spin correlation length is short~\cite{simkovic_two_particle}, is still debated, the relation of the pseudogap to ordered states is still under study~\cite{simkovic_pseudogap,shiwei_ground_state_stripe,shiwei_finite_T}, and many anomalous behaviors that manifest in materials remain unexplored.
This is especially true of dynamical quantities, which often display pronounced pseudogap phenomena in materials, but which are particularly challenging to compute numerically.

Cold atom-based quantum simulators offer an alternate approach to computational studies by providing pristine experimental realizations of the Hubbard model~\cite{hofstetter_hightc,tarruell_review}.
In such systems, spectroscopic and transport measurements grant direct access to dynamical quantities~\cite{waseem_arpes,waseem_bad_metal,martin_spin_transport,thywissen_conductivity,pritchard_polaron}.
Thermodynamic measurements have charted phase diagrams for correlated states of matter, in both bulk and lattice systems, that challenge numerical simulations~\cite{ku_unitary_thermodynamics,kohl_eos,salomon_eos,jochim_eos,navon_eos,munekazu_eos,kinast_eos,folling_eos,henning_pedestal}.
Widely tunable interaction strengths enable the controlled study of strong-coupling physics~\cite{tarruell_review}.
Further, quantum gas microscopy enables the detection and control of such systems at the level of single particles~\cite{gross_microscope}.
Using these techniques, recent measurements have highlighted the onset of high-order spin and charge correlations as the predicted pseudogap temperature is approached from above~\cite{bloch_pseudogap}.
So far, such experiments have been limited to comparatively high temperatures.
Recently, however, a several-fold reduction in achievable temperatures, enabled by improved cooling techniques, has produced samples deep within the regime where a pseudogap is expected~\cite{cooling_paper}.

\placefigure[!t]{f1}

In this work, we report the observation of a pseudogapped metal in the Hubbard model via thermodynamic and spectroscopic measurements of ultracold fermionic lithium in an optical lattice (Fig.~\ref{fig:fig1}a).
We find that the isothermal compressibility exhibits a peak versus doping at large interactions ($U/t\geq\Uv$) and low temperatures ($T/t<\Twidcold$), signaling an anomaly in the equation of state that separates an underdoped from an overdoped metal.
We further characterize these thermodynamic regimes by measuring an analog of electronic Raman scattering~\cite{devereaux_hackl_review} via lattice modulation spectroscopy.
In the underdoped regime, we find a depletion of low-energy excitations which is strongest in $B_1$ symmetry, indicating the formation of a pseudogap near the $(\pi,0),(0,\pi)$ points in the Brillouin zone, reminiscent of the `Fermi arcs' observed in cuprate superconductors~\cite{quantum_matter_high_Tc, norman_destruction_FS}.
Using low-frequency absorption data, we construct a pseudogap phase diagram which closely matches the thermodynamic phase diagram that is divided by the compressibility maximum (Fig.~\ref{fig:fig1}c).

These results establish the emergence of a pseudogapped metal at low temperatures, large interaction strengths, and small to intermediate dopings in the Hubbard model, which is both thermodynamically and dynamically distinct from the normal metal at large doping.
The compressibility maximum also suggests a critical tendency in the charge sector, whose origin and connection to the pseudogap is conjectured~\cite{sordi_widom,sordi_long,khatami_dca,wietek_sinha_forestalled} but not yet established (Fig.~\ref{fig:fig1}b).

\placefigure[!t]{f2}
\section*{Experimental setup}

Our experiments begin with an ultracold Fermi gas of lithium atoms in the two lowest hyperfine states, prepared with conventional laser cooling and evaporative cooling techniques.
Building on prior work~\cite{cooling_paper}, we then turn on an optical lattice and an optical box trap to divide this gas into a low-entropy, band-insulating core and a high-entropy, metallic reservoir.
The insulator is subsequently converted back into an ultracold Fermi gas by nearly extinguishing the lattice, and the hot atoms in the metallic reservoir removed with a magnetic gradient.
The remaining cold atoms in the Fermi gas are loaded into a second optical lattice with twice the density of sites, which produces Hubbard systems at very low temperatures (Methods~\ref{subsec:preparation},\ref{subsec:calibration}).

We use programmable optical potentials controlled by two digital micromirror devices (DMDs) to shape the density profile of the atomic cloud.
One DMD adds an anti-confining potential that compensates the potential of the lattice beams, producing a region of approximately uniform potential at the trap center, with RMS density variation below $3\%$.
The second DMD grants control over the central density by varying the intensity of a confining potential at the edge of the uniform region.
We tune and calibrate the Hubbard interaction $U$ and temperature $T$ of the system through standard methods (see Methods~\ref{subsec:preparation},\ref{subsec:calibration}). 
The result is a homogeneous sample of roughly $300$ sites with a tunable density $n$, temperature $T$, and Hubbard interaction $U$, embedded in a dilute reservoir of roughly $150$ sites. 
We measure the system by taking snapshots with a quantum gas microscope.

\section*{Compressibility maximum at low temperatures}

When the properties of a substance change upon cooling, it often manifests in the equation of state.
Classical fluids exemplify this behavior: their pressure-volume isotherms follow the ideal gas law $pV=Nk_BT$ at high temperatures, but flatten at low temperatures upon approaching the liquid-gas transition.
Quantum fluids exhibit analogous phenomena~\cite{kardar_particles}.
To detect changes in the physics of the Hubbard model at low temperatures, it is therefore desirable to measure the equation of state.

To do so, one must measure two conjugate thermodynamic quantities, such as pressure and volume.
In ultracold gases, the most convenient pair is usually density $n$ and chemical potential $\mu$~\cite{ku_unitary_thermodynamics,kohl_eos,salomon_eos,folling_eos,jochim_eos,navon_eos,munekazu_eos,henning_pedestal}.
For technical reasons, it is easier in our experiment to measure the isothermal compressibility $\kappa=\frac{\partial n}{\partial \mu}|_{T,U}$ than the global chemical potential $\mu$.
We thus choose to measure the $\kappa(n)$ equation of state, which is equivalent to the $n(\mu)$ equation of state.
Inspired by Ref.~\cite{henning_pedestal}, we measure the compressibility by applying a step $\Delta\mu$ in the chemical potential between the left and right halves of the system, and measuring the induced density difference $\Delta n\approx \kappa \Delta \mu$ (Fig.~\ref{fig:widom}a).
We give further details and discuss possible systematic errors in Methods~\ref{subsec:compressibility}.

To validate this procedure, we measure the compressibility at interactions of $U/t=\Uviii$ at a comparatively high temperature of $T/t=\Twidhot$. 
The measured compressibility (Fig.~\ref{fig:widom}b) is suppressed close to half-filling by proximity to the Mott insulator, and increases with doping in the metallic regime. 
This behavior reflects the smooth crossover from Mott insulator to Fermi liquid as a function of doping at high temperatures~\cite{gkkr}. 
We find reasonable agreement between the experiment and numerically exact determinant quantum Monte Carlo (DQMC) simulations (see Methods~\ref{subsubsec:dqmc}) at this elevated temperature (solid line).

However, at low temperatures of $T/t<\Twidcold$, the system becomes highly compressible at intermediate dopings of $\delta\sim[0.05-0.15]$.
The compressibility develops a peak above the high-temperature curve, displaying a pronounced local maximum. 
This maximum implies an inflection point in the $n(\mu)$ equation of state.
DQMC simulations at these low temperatures are infeasible due to the fermion sign problem (Methods~\ref{subsubsec:dqmc}).

The physical origin of the compressibility maximum is debated.
It was first predicted in cluster extensions of dynamical mean-field theory (DMFT)~\cite{sordi_long,sordi_widom,khatami_dca} as a precursor to a phase transition between a pseudogapped metal and a normal Fermi liquid.
The compressibility maximum is thus analogous to the inflection of isotherms in nearly-critical classical fluids, forming a Widom line above the predicted critical point~\cite{sordi_widom}.
This resembles measurements of cuprates, where thermodynamic quantities such as the specific heat show an anomaly near the pseudogap boundary~\cite{proust_taillefer_review,cuprate_criticality,loram_specific_heat}.

More recently, tensor network simulations of the Hubbard model have observed a similar compressibility maximum and attribute it to incipient charge stripe order which solidifies upon further cooling, producing an enhanced charge susceptibility $\chi_{ch}(q)$ at small but finite wavevectors $q\sim 2\pi\delta$~\cite{wietek_sinha_forestalled}.
Charge stripes are widely seen in ground-state, finite-size simulations of the Hubbard model, though they remain elusive in unbiased simulations in the thermodynamic limit, including at finite temperature~\cite{qin_computational_review,simkovic_two_particle,shiwei_finite_T}.
Because our system is $20$ sites in diameter, it is challenging to distinguish the uniform charge response $\kappa=\chi_{ch}(q=0)$ from the response at small but finite wavevectors $\chi_{ch}(q\sim 2\pi/20)$.
Thus it is possible that the observed compressibility enhancement indicates charge stripe order at finite $q$, rather than inflection in the thermodynamic-limit equation of state ($q=0$ response).
Alternatively, it is also possible that charge stripes affect the thermodynamic-limit equation of state.
Future experiments could provide decisive input to this issue by measuring the charge susceptibility $\chi_{ch}(q)$ versus $q$ at long wavelengths, and in larger systems.

We repeat the low-temperature measurement across a range of interaction strengths and plot the results in Fig.~\ref{fig:widom}c. We observe similar maxima in $\kappa(n)$ at $U/t=\Uv,\Uvi,\Uviii,$ and $\Uxi$, but not at $U/t=\Uiii$ or $\Uiv$ (Methods~\ref{subsubsec:widom_fit}).
The compressibility maximum thus demarcates an underdoped from an overdoped regime at low temperatures and strong interactions.
We infer that at low temperatures and strong interactions, two thermodynamically distinct regimes emerge in the doped Hubbard model.

\placefigure[!t]{f3}

\section*{Pseudogap and magnetic resonance via lattice modulation spectroscopy}

We now turn to the problem of characterizing these two regimes.
Spectroscopic measurements are particularly suited to this task, since dynamical properties often sharply reflect a change of state~\cite{simeoni_acoustic_widom,xu_widom}.
In both cluster DMFT~\cite{sordi_long,sordi_widom,ferrero_vbdmft,gull_dmft_comparative,ferrero_pseudogap_osm,werner_8_site} and in cuprates~\cite{arpes_on_cuprates_review,lee_wen_nagaosa_review,basov_electrodynamics,norman_pines_kallin_pseudogap}, passing from normal metal to pseudogap manifests as a selective loss of density of states near the antinodal $(\pi,0),(0,\pi)$ momenta.
This is strikingly realized in the `Fermi arc' observed in angle-resolved photoemission experiments on cuprates~\cite{arpes_on_cuprates_review} (schematized in Fig.~\ref{fig:U8_raman}a).
Hence we desire a spectroscopic probe of the antinodal momenta.

To this end, we realize a variant of lattice modulation spectroscopy~\cite{tillman_doublon,tillman_mott}, in which we measure the amount of heat generated by sinusoidal modulation of the lattice depth (Methods~\ref{subsec:spectroscopy}).
This measurement is analogous to electronic Raman scattering in solid state experiments, which is a powerful tool for studying the momentum-space structure of electron dynamics~\cite{devereaux_hackl_review}.
As a data efficient means of measuring heating, we convert the system back into a band insulator after modulation (Fig.~\ref{fig:U8_raman}a) and measure the number of singly-occupied sites.
In the language of Raman scattering, modulating the $x$ and $y$ lattice depths in (out of) phase grants access to the $A_1$ ($B_1$) response (Methods~\ref{subsec:spectroscopy}).
In a quasiparticle picture the $A_1$ and $B_1$ responses can be understood as coupling to different regions of the Brillouin zone~\cite{devereaux_hackl_review} (Fig.~\ref{fig:U8_raman}a).
The $B_1$ response probes the antinodal regions and is therefore of particular interest.
Indeed, measurements of the $B_{1g}$ Raman response in cuprates have found a suppression of low-energy weight with cooling, compatible with an antinodal gap~\cite{sacuto_cuprate_raman,sakai_cuprate_raman,sacuto_review}.
Cluster DMFT simulations of the $B_1$ Raman response in the Hubbard model also display a pseudogap~\cite{millis_raman_dmft}.

Prior measurements of the lattice modulation spectrum in optical lattice experiments have focused on a high-energy resonance $\omega\approx U$~\cite{tillman_mott,tillman_doublon}.
This feature is associated with doublon-hole production, and is well-understood theoretically, though there remain open questions on the possible substructure and dispersion of the resonance~\cite{bohrdt_exciton,tillman_doublon}.
To validate our spectroscopic procedure, in Fig.~\ref{fig:u8B1} we reproduce the $\omega\approx U$ feature in $B_1$ symmetry, and find good agreement with exact diagonalization simulations performed at the experimental parameters.
Our subsequent discussion focuses on the low-frequency response ($\omega/t<2.7$), which probes magnetic and quasiparticle excitations~\cite{devereaux_hackl_review}.

In Fig.~\ref{fig:U8_raman}b
we plot the $A_1$ and $B_1$ responses versus doping and frequency for $\delta=0-0.25, \omega/t=0-2.7$ at interaction strength $U/t=\Uviii$.
The data are symmetrized about $\omega=0$, since the response is an even function of frequency (Methods~\ref{subsec:spectroscopy}).
In both responses, at large doping, we observe a single peak near $\omega=0$, which we interpret as a Drude response compatible with normal metallic behavior~\cite{devereaux_hackl_review}.
As the doping is reduced across $\delta\approx 0.1$, however, both responses change.
The $A_1$ response at low frequencies smoothly drops towards zero, as the system is brought closer to the Mott insulator.
The $B_1$ response at low frequencies $\omega\lesssim 0.4t$ drops more sharply as $\delta\approx 0.1$ is crossed.
This behavior indicates a loss of metallicity which is non-uniform in the Brillouin zone, and hence a pseudogap.
Fig.~\ref{fig:U8_raman}c shows a direct comparison between the $A_1$ and $B_1$ response summed at frequencies $\omega/t=0.09,0.18$, approximating an integral over $\omega\lesssim0.2t$.
Recalling the location of the compressibility maximum, the data suggest a link between the thermodynamic changes evident in the compressibility and the formation of a pseudogap in the lattice modulation spectrum.

Upon crossing into the pseudogap, the $B_1$ response also develops a peak at $\omega/t=1.27(3)$.
At half-filling, this is known as a two-magnon peak, and can be understood as a process that produces pairs of magnons with opposite momenta~\cite{devereaux_hackl_review,cuprate_2mg_theory}.
The character of the resonance can be probed through quantum gas microscopy by measuring correlation functions versus time while the drive is applied (Fig.~\ref{fig:U8_raman}d, Fig.~\ref{fig:driven-corr}, and Methods~\ref{subsec:drivencorr}).
Fitting a sinusoid to this data grants access to both the amplitude and phase of the response.
We find that while density correlations show little response, spin correlations show a large response which lags the drive by $\sim\pi/2$ ($0.65(6)\pi$ at $\omega/t=1.44$), confirming the magnetic nature of the resonance (Fig.~\ref{fig:U8_raman}d).
In units of the superexchange energy $J=4t^2/U$, the two-magnon peak frequency is $\omega/J=2.24(5)$.
This is lower than the value $\omega/J=2.7-3$ commonly found in the Raman scattering literature on cuprates~\cite{devereaux_hackl_review}.
However, increasing $U/t$ to $\Uxi$ shifts the frequency to $\omega/J=2.6(2)$ (Fig.~\ref{fig:raman_vs_U}c, Extended Data Table~\ref{tab:spectroscopy}), closer to the canonical value.
This suggests an effective reduction of the magnetic exchange energy below $4t^2/U$ due to the finite value of $U/t$.

At a doping $\delta=0.09$, between the pseudogapped and normal metallic regimes, the $B_1$ response becomes flat over an order of magnitude in frequency, from $\omega/t=0.13$ to $\omega/t=1.44$ (Fig.~\ref{fig:U8_raman}c).
Given that this density lies close to the compressibility maximum (Fig.~\ref{fig:widom}), it is possible that this frequency-independent behavior reflects some type of criticality~\cite{subir_qpt,cuprate_criticality}.

To further probe the link between the pseudogap and the compressibility maximum, we measure the pseudogap versus interaction strengths.
In Fig.~\ref{fig:raman_vs_U}a,b, we plot the low-frequency ($\omega\lesssim 0.2t$) $A_1$ and $B_1$ responses.
A region of suppressed response is apparent at low doping in the $B_1$ response at interactions $U/t\geq \Uv$.
By fitting each curve versus doping to a sigmoid function, we extract a characteristic doping $\delta_\star$ for the onset of the pseudogap at each $U$ (Methods~\ref{subsec:raman_fit}).
This establishes a pseudogap phase boundary in the $\delta-U$ plane, which is overlaid on Fig.~\ref{fig:raman_vs_U}b.
The $A_1$ response, by contrast, drops more smoothly as the doping is reduced towards half-filling at large interactions, showing a less pronounced loss of absorption in the underdoped regime.
At large interactions and finite doping, the low-frequency $A_1$ response qualitatively resembles the compressibility (Fig.~\ref{fig:widom}c), suggesting a connection to the thermodynamic changes signaled by the compressibility maximum.
At low interactions $U/t=\Uiii$, both responses show little structure versus doping.

We also find that as we vary $U$, the appearance of the pseudogap (i.e. loss of low-frequency $B_1$ absorption) is associated with the presence of the two-magnon peak at half-filling.
As shown in Fig.~\ref{fig:raman_vs_U}c, the energy of this peak varies non-monotonically with interaction strength.
At $U/t=\Uiii$, where we cannot resolve a pseudogap in the $B_1$ response, we also cannot distinguish the two-magnon resonance from a Drude peak.

\placefigure[!t]{f4}

\section*{Discussion}

One of the main open questions about the Hubbard pseudogap concerns the similarities and differences between the physics at weak and strong interaction strengths.
At weak interactions, a type of pseudogap is known to occur in the Hubbard model due to the scattering of quasiparticles off of antiferromagnetic spin fluctuations~\cite{vilk_tremblay_1997,schafer_footprints}.
However, this mechanism is only effective when the antiferromagnetic correlation length is sufficiently long.
Both in hole-doped cuprates~\cite{kastner_cuprate_review} and the present work (see Fig.~\ref{fig:corr-length}), the correlation length is short, suggesting physics beyond the weak-coupling theory is relevant.
Precisely identifying this physics remains an open problem.

In our measurements, signs of the pseudogap are restricted to strong interaction strengths $U/t\geq\Uv$, suggesting a qualitative distinction between the weak- and strong-coupling doped regimes.
At the lower interaction strengths $U/t=\Uiii,\Uiv$ explored in this work, we find neither the anomalous thermodynamics nor the rapid changes versus doping in the lattice modulation spectrum that characterize the underdoped regime at strong coupling.
In this respect, our data are in qualitative agreement with cluster extensions of DMFT~\cite{sordi_long} in which the pseudogap  is associated with Mott physics, restricting its dynamic and thermodynamic signatures to strong coupling.
Fermi arc-like structures, however, are known to appear in the one-particle spectral function at both weak~\cite{simkovic_pseudogap,vilk_tremblay_1997,schafer_footprints} and strong coupling~\cite{simkovic_pseudogap}.

This raises the intriguing possibility that criteria beyond a tendency towards Fermi arc formation are important in understanding the strong-coupling pseudogap, and that the behaviors identified in this work are among such criteria.
Notably, both experimental quantities are two-particle correlation functions.
In the weak-coupling pseudogap, two-particle quantities like spin correlations remain Fermi liquid-like even as the one-particle function becomes pseudogapped~\cite{vilk_tremblay_1997}.
Indeed, despite the prominence of the Fermi arc in the phenomenology of the cuprate pseudogap, it is worth recalling that many of its anomalous features concern two-particle quantities like the spin susceptibility~\cite{alloul_1989}.
Understanding possible distinctions in pseudogap behavior between one- vs two-particle quantities is thus an important problem.
Understanding the effects of temperature and of deviations from the `vanilla' Hubbard model, both of which can affect the one-particle pseudogap~\cite{simkovic_pseudogap,vilk_tremblay_1997}, in our measurements is also an important future direction (see Methods~\ref{subsubsec:bandgap_spectroscopy},~\ref{subsubsec:T_vs_U}).

In our data, the compressibility maximum implies a link between the pseudogap and charge correlations that onset at low temperatures.
Such a link has been conjectured, but not definitively established, in the numerical literature~\cite{simkovic_pseudogap,shiwei_ground_state_stripe,shiwei_finite_T,simkovic_two_particle}, and pertains to the long-debated relation between the pseudogap and stripe order.
Recently, state-of-the-art diagrammatic Monte Carlo and constrained path auxiliary field quantum Monte Carlo simulations find that the one-particle pseudogap evolves into stripe order at zero temperature, though the finite-temperature transition to charge stripe order is not yet understood~\cite{simkovic_pseudogap,shiwei_ground_state_stripe,shiwei_finite_T,simkovic_two_particle}.
Tensor network~\cite{wietek_sinha_forestalled} and cluster DMFT~\cite{sordi_long,sordi_widom,khatami_dca} simulations, taken together, suggest a somewhat different picture, in which charge stripes emerge from a range of highly compressible densities generated by the passage from pseudogap to normal metal.
This recalls theoretical proposals on cuprates, in which a hypothesized microscopic tendency to phase separation gives rise to one or several `intertwined' orders~\cite{kivelson_phase_separation,intertwined_review}.

Precisely understanding the origin and nature of the compressibility maximum would thus inform a set of centrally important questions on the Hubbard model -- for instance, whether the pseudogap is better understood as a precursor to an ordered state, or a distinct metallic state which generates ordered states at its boundary.
Such questions are at the limits of modern computational techniques, but can now also be probed by experiments.

\section*{Outlook}

This work experimentally demonstrates the existence of a pseudogapped metal at low temperatures and strong interaction strengths in the underdoped regime of the square lattice Hubbard model.
The $A_1$ and $B_1$ lattice modulation responses show a deviation from conventional metallic behavior, a hallmark of the pseudogap.
Departing the pseudogap results in a peak in the compressibility, marking the pseudogap metal as thermodynamically distinct from the normal metal at large doping.

An important direction for future work is to understand the relationship between charge stripe order and the pseudogap.
As discussed above, the compressibility maximum suggests a critical tendency in the charge sector which is linked to the pseudogap, but whose precise nature remains unclear.
Directly measuring $\chi_{ch}(q)$ versus $q$, by replacing the step potential with a sinusoidal potential, would help resolve this issue.

Realizing the pseudogap in a Hubbard quantum simulator also opens two complementary fronts of research: experiments can now study how closely the Hubbard pseudogap mimics the cuprate pseudogap, and also explore entirely new approaches to understand the pseudogap.
Regarding the former goal, we note that while the cuprate pseudogap is extensively documented as a related set of phenomena across many different experimental probes~\cite{norman_pines_kallin_pseudogap,lee_wen_nagaosa_review,proust_taillefer_review}, the same is not yet true of the Hubbard pseudogap.
Additionally, the cuprate pseudogap is typically studied in the $\delta-T$ plane, rather than the $\delta-U$ plane used in this work.
Improved thermometry and further measurements of the pseudogap's physical properties are thus important in this context.

Studying the pseudogap versus new experimental tuning parameters, accessible in a quantum simulator, may reveal more of its physics.
Our data versus interaction strength are a first step in this direction.
Measurements versus band structure could provide further insights.
It is proposed, for instance, that working on magnetically frustrated lattices~\cite{muqing_triangle,schauss_mott} can test the relative importance of Mott physics and antiferromagnetism to the pseudogap~\cite{downey_doped_triangle}.

More advanced analyses of snapshot data in the pseudogap regime could also bring new insights.
In general, measurements of higher-order correlation functions are a unique experimental capability of quantum simulators, allowing one to detect phenomena which remain hidden in more conventional types of response functions~\cite{chiu_string,hilker_hidden}.
Image classification techniques, for example, have been used to construct new observables that show signs of novel physics~\cite{ehsan_structural_snapshot,annabelle_correlator}.
Such measurements are particularly useful in regimes that, like those in this work, challenge classical numerical methods.
Pairing snapshot analysis with the appropriate modulation protocol, as we begin to explore in Fig.~\ref{fig:U8_raman}d, could be a powerful tool for understanding the nature of different spectral features.
For example, such measurements could reveal the origin of the flat $B_1$ response in Fig.~\ref{fig:U8_raman}b, and any connections to criticality.

More broadly, this work signals the arrival of quantum simulators at the frontier of knowledge on the basic model of correlated electrons, namely the square lattice Hubbard model, and demonstrates the capacity of such experiments to have productive input into open problems.
The methods developed in this work, including the production of large, homogeneous samples at low temperatures and the measurement of low-frequency response functions, will likely be important in future studies of the Hubbard phase diagram, including those indicated above.
With open questions remaining on the physics of low-temperature phenomena like stripes and $d$-wave superconductivity, we believe that there is much to learn from quantum simulations of the Hubbard model.

\subsection*{Acknowledgments}
We thank Chunhan Feng, Antoine Georges, Dominik Kiese, Olivier Parcollet, and Shiwei Zhang for extensive discussions and sharing of preliminary numerical data.
We thank Muqing Xu for early experimental contributions.
We thank Radu Andrei, Tizian Blatz, Annabelle Bohrdt, Petar Bojovi\'c, Pietro Bonetti, Jonathan Curtis, Eugene Demler, Tom Devereaux, Fabian Grusdt, Ehsan Khatami, Steve Kivelson, David LeBoeuf, Ivan Morera, Alexander Nikolaenko, Subir Sachdev, Aritra Sinha, Giovanni Sordi, Andr\'e-Marie Tremblay, and Alexander Wietek for insightful discussions.
We acknowledge support from the Gordon and Betty Moore Foundation, Grant No.~GBMF-11521;
National Science Foundation (NSF) Grants Nos.~S6277 and OAC-2118310;
ONR Grant No.~N00014-18-1-2863;
the Department of Energy, QSA Lawrence Berkeley Lab award No.~DE-AC02-05CH11231;
QuEra grant No.~A57912; ARO/AFOSR/ONR DURIP Grant No. W911NF-20-1-0163; 
ARO ELQ Award No.~W911NF2320219;
the NSF Graduate Research Fellowship Program (L.H.K. and A.K.);
the AWS Generation Q Fund at the Harvard Quantum Initiative (Y.G.);
the Swiss National Science
Foundation (M.L.);
the Harvard Quantum Initiative Graduate Fellowship (A.D.D.);
the Intelligence Community Postdoctoral Research Fellowship Program at Harvard administered by Oak Ridge Institute for Science and Education (ORISE) through an interagency agreement between the U.S. Department of Energy and the Office of the Director of National Intelligence (ODNI) (A.W.Y.).

\subsection*{Author Contributions}
L.H.K., A.K., Y.G., A.D.D., M.L., and A.W.Y. performed the experiment and analyzed the data.
L.H.K. and A.K. performed the numerical simulations.
M.G. supervised the study.
All authors contributed to the interpretation of the results and production of the manuscript.

\subsection*{Competing Interests}
M.G. is co-founder, shareholder, and consultant of Quera Computing.

\begin{dfigure*}{f1}
\centering
\noindent
\includegraphics[width=6.50in]{"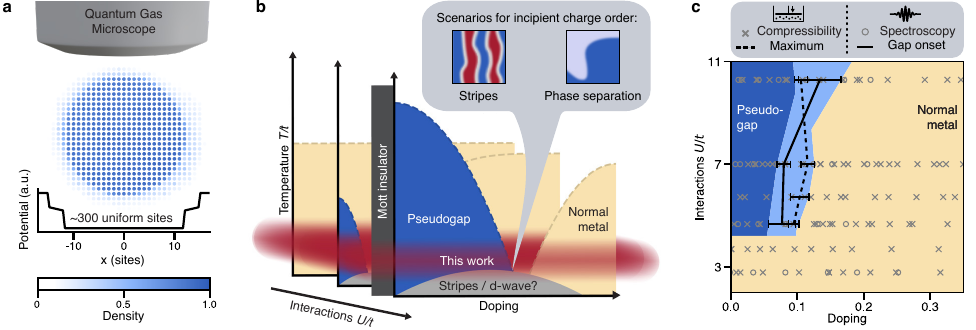"}
    \caption{\textbf{Hubbard pseudogap in a quantum gas microscope.}
    (\textbf{a}) Schematic of experiment. We produce samples of ultracold lithium atoms in a Hubbard-regime optical lattice, controlling the local potential to obtain roughly $300$ uniform sites embedded in a bath of roughly $150$ sites. We measure the system by taking snapshots with a quantum gas microscope. The color plot is an average of many such snapshots.
    (\textbf{b}) Schematic phase diagram of the Hubbard model versus temperature, interaction strength, and doping, showing Fermi liquid, pseudogap, and Mott insulating regions, together with conjectured stripe or $d$-wave superconducting regions. The experimental regime is indicated with a red shaded region. 
    (\textbf{c}) Experimental phase diagram. The solid line indicates the onset of the spectroscopic pseudogap and the dashed line is the compressibility maximum (both defined in the text).
    The light blue region is subtended by these two lines and their errorbars. The pseudogap and normal metallic regions are indicated on either side of this line.
    }
    \label{fig:fig1}
\end{dfigure*}

\begin{dfigure*}{f2}
    \centering
    \noindent
    \includegraphics[width=6.33in]{"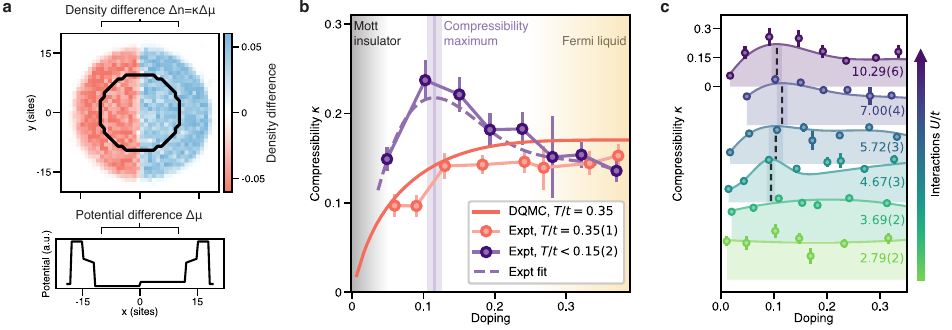"}
    \caption{
    \textbf{Compressibility maximum at low temperatures.}
    (\textbf{a}) Experimental protocol to measure compressibility. A chemical potential step $\Delta\mu$ is applied to the system (lower panel). The induced density difference $\Delta n\approx \kappa\Delta \mu$ senses the isothermal compressibility $\kappa=\frac{\partial n}{\partial \mu}|_{T,U}$. The upper panel shows the difference between the average of a snapshot ensemble taken with $\Delta\mu>0$ and an ensemble taken with $\Delta\mu<0$. The typical density difference induced by $\Delta\mu$ is $0.04$. The black circle indicates the homogeneous region in the trap center. (\textbf{b}) Compressibility at $U/t=\Uviii$. The solid red curve is exact numerical data obtained at high temperature ($T/t=\Twidhotn$) using DQMC. The red points are experimental data at high temperature ($T/t=\Twidhot$), and show reasonable agreement with DQMC.
    The purple points are low-temperature ($T/t<\Twidcold$) experiment data, and show an enhanced compressibility over high temperatures at doping $\delta=0.05-0.15$.
    This maximum in the compressibility implies an inflection point in the $n(\mu)$ equation of state, providing a thermodynamic distinction between the underdoped and overdoped metal.
    The dashed purple line is a fit to the experimental data and is used to extract the doping of the compressibility maximum (vertical purple line; see Methods~\ref{subsubsec:widom_fit}).
    (\textbf{c}) Compressibility versus doping and interactions at low temperatures.
    The compressibility maximum (vertical dashed lines) is seen to persist at strong interactions $U/t\geq \Uv$, but vanishes at weak interactions $U/t=\Uiii,\Uiv$ (Methods~\ref{subsubsec:widom_fit}).
    }
    \label{fig:widom}
\end{dfigure*}

\begin{dfigure*}{f3}
    \centering
    \noindent
    \includegraphics[width=7.11in]{"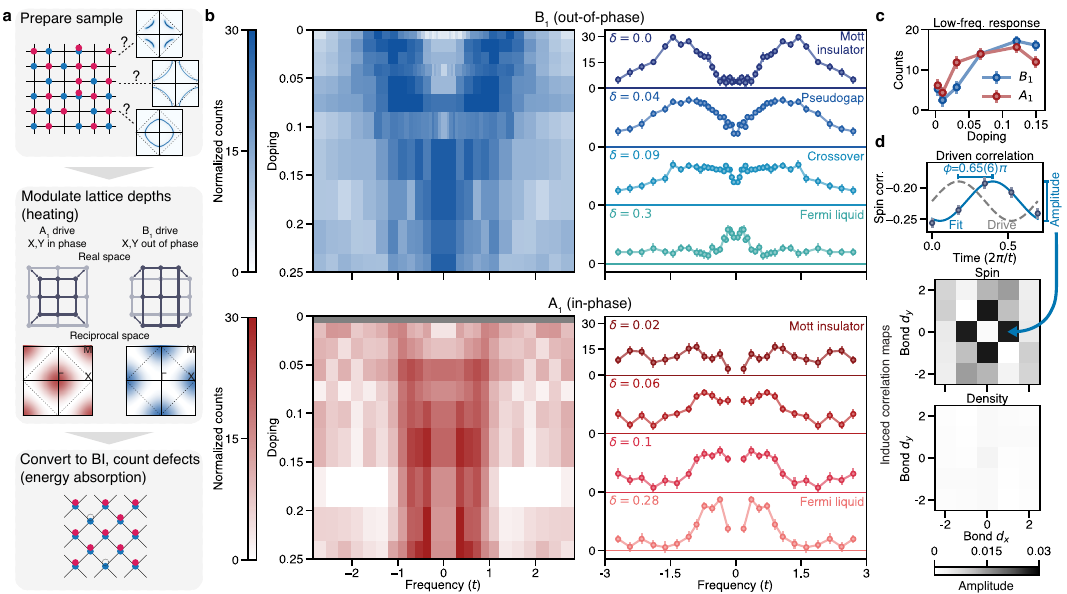"}
    \caption{\textbf{Pseudogap and magnetic resonance in lattice modulation spectroscopy at $U/t=\Uviii$.}
     (\textbf{a}) Spectroscopic protocol. Modulating the $x$ and $y$ lattices in- or out-of-phase couples to different regions of the Brillouin zone. The $\Gamma=(0,0),M=(\pi,\pi),$ and $X=(\pi,0)$ points are indicated for clarity. The response is measured by converting the system into a band insulator and measuring the number of defects, a proxy for energy absorption.
     (\textbf{b}) Spectra versus frequencies $\omega/t<2.7$ and dopings $\delta=0-0.25$ in $A_1$ and $B_1$ symmetry. On the underdoped side of the compressibility maximum (Fig.~\ref{fig:widom}), the $B_1$ spectrum displays a loss of weight at low energies, suggesting a spectral gap close to the $X$ point. The $B_1$ response also develops a resonance which connects to the two-magnon peak of the Mott insulator at $\delta=0$. The $A_1$ response, by contrast, changes more smoothly. At large doping, both $A_1$ and $B_1$ responses exhibit a Drude-like feature, suggesting metallic behavior.
     (\textbf{c}) $A_1$ and $B_1$ responses averaged at $\omega/t=0.09,0.18$ versus doping, showing the differential suppression in the two responses in the underdoped regime.
     (\textbf{d})
     Time trace of nearest-neighbor spin correlations while the drive is applied near the two-magnon peak ($\delta=0$, $\omega/t=1.44$).
     A sinusoidal fit yields the amplitude and phase of the response.
     The $\phi\sim \pi/2$ phase lag indicates a resonant response.
     Plotting a map of these amplitudes for additional correlations, we see a large response in spin correlations but not density correlations, confirming the magnetic nature of the resonance.
    }
    \label{fig:U8_raman}
\end{dfigure*}

\begin{dfigure}{f4}
    \centering
    \noindent
    \includegraphics[width=3.4in]{"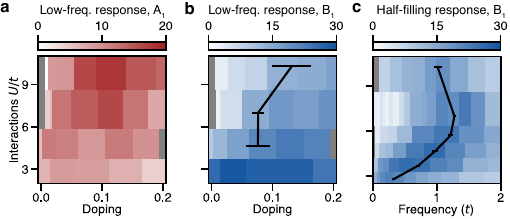"}
    \caption{\textbf{Spectroscopic phase diagram.}
    (\textbf{a}) and (\textbf{b})
    Summed response at $\omega/t=0.09,0.18$ for $A_1$ and $B_1$ drives, approximating an integral for $\omega\lesssim 0.2t$.
    The pseudogapped region can be identified by the suppressed response at low doping and large interactions in the $B_1$ response. The black line marks $\delta_{\star}$, the fitted characteristic doping for the pseudogap onset at each $U/t$ (Methods~\ref{subsec:raman_fit}). 
    (\textbf{c})
    Half-filled (i.e. $\delta=0$) $B_1$ response as a function of $U$ and frequency.
    The two-magnon peak frequency (black line, Methods~\ref{subsec:2mp_fit}) is non-monotonic in $U/t$.
    At the lowest $U/t$ studied, we cannot distinguish a two-magnon peak from a Drude-like response.
    }
    \label{fig:raman_vs_U}
\end{dfigure}

\bibliography{paper.bib}

\newpage

\setcounter{figure}{0}

\renewcommand{\figurename}{Extended Data Figure}
\renewcommand{\thefigure}{S\arabic{figure}} 
\renewcommand{\tablename}{Extended Data Table}
\renewcommand{\thetable}{\arabic{table}} 

\clearpage

\section*{Methods}

\subsection{Sample preparation}\label{subsec:preparation}
We outline the experimental sequence in Fig.~\ref{fig:rampfig}a.
We first prepare an ultracold Fermi gas using standard laser and evaporative cooling techniques~\cite{daniel_mott}.
Subsequent steps follow a modified version of the cooling protocol described in Ref.~\cite{cooling_paper}.
First, we form a band insulator (BI) from the central, degenerate region of the Fermi gas by loading it into an optical lattice and a programmable potential on one of the DMDs (DMD0).
The lattice is formed by two nearly orthogonal interfering lattice beams, which we call $X$ and $Y$~\cite{muqing_triangle}. We call this lattice the \textit{sparse lattice}.
The DMD potential is a `volcano' pattern (Fig.~\ref{fig:rampfig}a), in which the central $12$-site radius region is dark, a sharp step is imposed at radius $12$, and the potential drops to zero with a parabolic form outside radius $12$.
Ramping up this potential during lattice loading serves to first enhance the central density during BI formation and then separate the BI from the metallic reservoir as the DMD is ramped up~\cite{cooling_paper}.

With the sparse lattice at its full depth, we then transfer the BI to a different potential on the other DMD (DMD1).
This `countersink' potential is dark out to radius $12$, has a sharp step at radius $12$,  slopes upward out to radius $18$, and has a sharp wall at radius $18$.
Empirically, this inward-sloping wall improves the adiabaticity of the eventual flow of atoms out of the central dark region, which is needed to produce doped systems~\cite{cooling_paper}.
Handing off the BI from DMD0 to DMD1 also allows us to swap the pattern on DMD0 to the anticonfining potential described in Section~\ref{subsubsec:potential}, while maintaining a strong potential barrier between the BI and the hot reservoir atoms, for thermal isolation.

After the handoff from DMD0 to DMD1, we ramp off the sparse lattice, converting the BI into an ultracold Fermi gas.
We then remove the hot reservoir atoms with a magnetic gradient (not shown in Fig.~\ref{fig:rampfig}).
This `cleanup' stage (Fig.~\ref{fig:rampfig}a) eliminates the need for thermally isolating walls, making it possible to use weaker walls at later stages in the sequence.

After removing the hot reservoir atoms, we load the Fermi gas inside the DMD1 potential into a second optical lattice, which we call the \textit{dense lattice}, formed by the above-mentioned $Y$ beam and an additional $\bar{X}$ beam, which co-propagates with $X$ but does not interfere with $Y$~\cite{cooling_paper}.
At the same time, we turn on an anticonfining potential on DMD0, which compensates the roughly harmonic potential of the dense lattice (Section~\ref{subsubsec:potential}, Fig.~\ref{fig:rampfig}a).
The DMD1 potential is decreased during this loading stage.
Its final value decides what density is realized in the central uniform region: lower values allow more atoms to flow out of the center, reducing the central density.
As in prior work~\cite{cooling_paper} the magnetic bias field is ramped at the start of this ramp to its final value.
This value controls $U/t$ through control of the $s$-wave scattering length~\cite{tarruell_review}.
This final ramp produces a cold, homogeneous Hubbard system at a tunable density and interaction strength (Fig.~\ref{fig:rampfig}c).
We vary the temperature by holding the atoms for a variable duration at this stage of the sequence.

To measure the compressibility (Section~\ref{subsec:compressibility}) or spin correlations (Section~\ref{subsec:calibration}), we image the system at this point using a quantum gas microscope.
To measure the lattice modulation response, at this point in the sequence we modulate the lattice depths, then reverse the steps of the sequence to convert the system back into a BI (Fig.~\ref{fig:rampfig}a), which we measure with the quantum gas microscope.

\placefigure[!ht]{SA}

Converting the system back into a BI also provides a data-efficient means of characterizing and optimizing the experimental sequence.
Since the number of singly-occupied sites in the insulator is a proxy for the insulator's entropy~\cite{cooling_paper}, reversing the sequence to produce a BI after each stage allows a semi-quantitative measurement of the entropy added by each stage.
These values are shown in Fig.~\ref{fig:rampfig}b.
The various ramp times and DMD1 setpoints used in the sequence are optimized to minimize these values, while maximizing the total number of atoms retained in the BI.

\subsection{Sample preparation}

\subsubsection{Controlling the local potential}\label{subsubsec:potential}

The behaviors of interest in this work occur at low energy scales, are very sensitive to doping, and possibly involve large length scales.
As a result, it is critically important to prepare large, homogeneous samples, which in turn requires careful control of the local potential in the lattice.
The intrinsic harmonic confinement and disorder in the optical lattice are too large to be compatible with the above requirements.
To remedy this, we characterize the lattice potential \textit{in situ} using the atoms, and apply a correction that is projected through the microscope objective, and controlled using a digital micromirror device (DMD).

To apply the appropriate DMD potentials, we perform closed loop feedback with a camera.
Although the DMDs only allow for binarized control of intensity in an image plane of the atoms, the aperture of the projection system is chosen such that the point spread function spans multiple pixels.
We can therefore more finely control the amplitude of the illumination using spatial dithering.
Based on measurements with the camera, we achieve RMS errors below $1\%$ of the maximum applied DMD potential.
However, measurements with the atoms indicate residual RMS errors on the scale of $5\%$ of the applied potential, suggesting that there are differences between the camera optics and the projection optics leading to the atoms.

To characterize the lattice potential, we perform measurements with the same lattice parameters as in the rest of the work, but in a magnetic field of $554.64$G.
The resulting interactions of $U/t=\Ucal$ are sufficiently low that one can perform accurate simulations at low temperatures using DQMC.
We can therefore compare measurements of the density of singly occupied sites (via parity projected images of the atoms) to DQMC results and invert the equation of state to obtain a local potential.
Note that the local potential changes rapidly as a function of position in comparison to the value of the tunneling energy itself, and so we assume that the tunneling energy is constant as a function of position throughout this work.

The lattice potential is homogenized using an iterative procedure in which we collect $\sim 200$ experimental trials to characterize the potential, apply the appropriate correction using a DMD, and then perform additional rounds of measurements with $\sim 400$ trials to characterize any residual errors.
Typically, we find that a single additional measurement of residual errors is sufficient to reduce the disorder in the lattice to an RMS variation of $\lesssim0.1~t$.
The disorder in the lattice is stable on the scale of $\sim$ 1 week, beyond which slow lattice beam-pointing drifts cause a change in the shape of the lattice potential. Thus approximately once a week, we manually realign the lattice beams by maximizing the retro-reflected power that is coupled back into the lattice optical fibers, and repeat the above flattening procedure.

The alignment between the optical lattice and the DMD potential can change by a fraction of a lattice site on the timescale of several hours.
This generates gradients on the scale of $0.01~t/a$, where $a$ is the lattice spacing.
To correct this, we perform an automated correction every $2-5$ hours which involves measuring the local potential using 40 experimental trials, fitting a gradient, and regenerating the applied DMD potentials with the appropriate offset.
With the above automated procedure in place, the experiment can run for more than a week without human intervention while maintaining low temperatures, and gradients below $\sim0.005 t/a$, allowing us to gather the large number of experimental trials required for this work.
The corresponding peak-to-peak density variation in the region of interest in this work (with a radius of 10 lattice sites) is typically below $\sim\pm3\%$, depending on the precise Hubbard parameters.

\subsection{Data Collection}
\label{subsec:collection}
The data in this work comprise four datasets taken in the manner described above, in which the experiment was operated continuously for one to several weeks.
We refer to these as DS0, DS1, DS2, and DS3.
Datasets DS0 and DS3 are compressibility measurements, while datasets DS1 and DS2 are spectroscopy.
Datasets DS2 and DS3 use the doublon-resolved imaging described in Section~\ref{subsec:imaging}.
DS0, DS1, DS2, and DS3 respectively contain $23578,35552,23120,$ and $26892$ experimental repetitions, excluding data used in temperature and imaging calibration (Sections~\ref{subsec:imaging},~\ref{subsec:calibration}).
All of the data has interleaved thermometry data in which we measure the singles density and spin correlations at $\delta\approx 0.1$, $U/t=\Uviii$, as a check that the sample remains cold.

We additionally make use of three other small datasets, which we denote DSTherm, DSConv, and DSLat.
These were respectively taken during DS0, between DS2 and DS3, and between DS2 and DS3, and are given different names purely for notational clarity.
We make use of them, respectively, in Sections~\ref{subsubsec:T_vs_U},~\ref{subsec:analysis} and~\ref{subsubsec:bandgap_spectroscopy}.

\subsection{Compressibility measurement}\label{subsec:compressibility}
We measure the compressibility by applying a potential step $\Delta\mu$ to the atoms and measuring the induced density difference $\Delta n\approx \kappa \Delta\mu$.
The potential step is applied by modifying the DMD correction potential (see Section~\ref{subsubsec:potential}).
This method mimics the compressibility measurement in Ref.~\cite{henning_pedestal}.
The choice of a step potential, in our case, maximizes signal-to-noise for a fixed strength of the perturbing potential, by placing all sites at a maximal potential difference from the mean potential.
Within the local density approximation (in which the density of a site is set by its potential, with no effects from neighboring sites), any potential that puts half the sites at $+\Delta\mu/2$ and half at $-\Delta\mu/2$ has this property.
The choice of a step further maximizes the spatial extent of contiguous regions of uniform potential, to come as close as possible to measuring the uniform compressibility, rather than compressibility at finite wavelength.

Being a finite-difference method, this measurement really senses an average of the compressibility over a range $\Delta\mu$ of chemical potentials, as
\begin{equation*}
\kappa_{\rm{avg}}=\frac{n(\mu_0+\Delta\mu)-n(\mu_0)}{\Delta\mu}=\frac{1}{\Delta\mu}\int_{\mu_0}^{\mu_0+\Delta\mu}d\mu' \kappa(\mu').
\end{equation*}
So long as the compressibility is smooth on a scale of $\Delta\mu$, $\kappa_{\rm{avg}}$ should thus provide an accurate measurement.
If the compressibility has a peak with width $\lesssim \Delta\mu$, $\kappa_{\rm{avg}}$ will underestimate its height and broaden it to a width $~\Delta\mu$.
Thus the finite-difference nature of the measurement cannot spuriously produce a peak in the compressibility, but can broaden or distort peaks.

We also note that disorder in the local potential can lead to broadening of the compressibility peak measured in the experiment.
Within the local density approximation, this can be understood as taking an average of $\kappa(n)$ over the distribution of the local potential $n$.
Similarly to the average associated with finite-difference, this averaging will only substantially affect the measured $\kappa$ if $\kappa(\mu)$ varies rapidly on the scale of the disorder.
This averaging effect will also lead to broadening and lowering of peaks in the compressibility, but cannot spuriously produce a peak.

To gain robustness against disorder and drifts in the lattice potential (see Section~\ref{subsubsec:potential}), we perform this measurement in a differential way.
Specifically, half of the experimental repetitions use a positive step $\Delta\mu$, while the other half use a negative step $-\Delta\mu$.
The experimental compressibility is then computed as
\begin{equation*}
    \kappa=\frac{\Delta n(\Delta\mu)-\Delta n(-\Delta \mu)}{2\Delta\mu}.
\end{equation*}
The advantage of this procedure over a non-differential measurement is that the sensed density difference is predominantly due to the difference of two DMD potentials, rather than the combination of the DMD potential and lattice potential.
In a non-differential measurement, residual disorder in the local potential, whether due to the finite precision of the correction procedure described in Section~\ref{subsubsec:potential} or to drifts in the potential, effectively contaminates the step $\Delta\mu$.
In a differential measurement, such contamination is common to the positive and negative step, and drops out of the computed compressibility at linear order.
The repetitions using positive and negative $\Delta\mu$ are interleaved, so that slow drifts remain common to the two datasets.

The density difference $\Delta n$ is measured as a difference of a `left' and `right' density, $n_{L,R}$.
These densities are averaged within semi-circular regions on the left and right halves of the central radius-$10$ region (see Section~\ref{subsubsec:potential}).
We exclude the outermost ring of sites from this average, because we believe that the potential in these regions may deviate from its targeted value due to the limited resolution of our imaging and projection system.
This would reduce the signal to noise ratio of the compressibility measurement.

We calibrate $\Delta\mu$ by measuring the compressibility of a weakly interacting ($U/t=\Ucal$), heavily doped ($\delta\sim 0.5$) system.
This regime is straightforward to simulate using DQMC.
We thus obtain $\Delta\mu$ by requiring the experimental compressibility match DQMC at this interaction and doping.
This stepsize calibration data is interleaved with the data in datasets DS0 and DS3.
In DS0 we use a $\Delta\mu$ close to $0.2t$, while in DS3 we use a $\Delta\mu$ close to $0.3t$.
This choice is motivated both by numerical data~\cite{sordi_long,wietek_sinha_forestalled} and by preliminary experimental data taken before DS0.

We compute the $U/t=\Ucal$ DQMC compressibility at a temperature $T/t=0.3$. We find this is cold enough that the compressibility is essentially temperature-independent (as expected in a Fermi liquid below the degeneracy temperature) but hot enough that finite-size effects may be avoided with modest computational effort (see Section~\ref{subsubsec:dqmc}).
We note that deviations of the experimental calibration data from the condition simulated in DQMC will cause a systematic error on all compressibility data in the form of a global scale factor.
In particular, the experimental compressibility is computed as $\kappa_{\exp}=(\Delta n/\Delta n_{\mathrm{cal}})\times\kappa_{\mathrm{DQMC}}$, where $\Delta n_{\rm{cal}}$ is measured at $U/t=\Ucal$.
While the absolute accuracy of $\Delta n,\Delta n_{\rm{cal}}$ is limited by the imaging fidelity to a $\sim 1\%$-level (Section~\ref{subsec:imaging}), the accuracy of $\kappa_{\rm{DQMC}}$ to the experiment is limited by various calibration procedures (Section~\ref{subsec:calibration}).
Such deviations between simulation and calibration data are thus likely a leading source of systematic error.
They may explain, for example, why in Fig.~\ref{fig:widom}b, the $T/t=0.35$ DQMC curve lies systematically above the experiment.

In DS0, we observe statistically significant drifts in the calibrated stepsize $\Delta\mu$ over the course of data collection (Fig.~\ref{fig:combination_fig}a).
This drift is present in the inferred $\Delta\mu$ despite the fact that the programmed and measured stepsize on the DMD camera (see Section~\ref{subsubsec:potential}) is constant over this period.
We believe this effect is due to cropping of the DMD light close to the Fourier plane of our imaging system.
This cropping could produce an additional potential on the atoms, which would not drop out due to the differential measurement because it interferes coherently with the step potential, rather than adding linearly.
To avoid this effect in DS3, we blurred the pattern on the DMD to minimize cropping in the Fourier plane.
In DS3, the step size is constant over time to within our statistical uncertainty (Fig.~\ref{fig:combination_fig}a).

Due to this difference between DS0 and DS3, in DS0 we use a $\Delta\mu$ calibration that varies on the timescale of a day, so that the density difference data $\Delta n$ is only ever divided by a $\Delta\mu$ taken from the same span of time.
In DS3 we use a single value of $\Delta\mu$ for all the data, obtained by averaging all the calibration data together.

The curves shown in Fig.~\ref{fig:widom} are obtained by averaging DS0 and DS3 together.
This averaging procedure is described in Section~\ref{subsubsec:combination}.

\subsection{Lattice modulation spectroscopy}\label{subsec:spectroscopy}

The spectroscopic measurements in this work probe the imaginary part of a current-current susceptibility, which is similar to quantities probed via electronic Raman scattering in real materials~\cite{devereaux_hackl_review}.
Our measurement involves sinusoidal modulation of the lattice potential, followed by a measurement of the associated heating.
The time-dependent part of the perturbing Hamiltonian is:
\begin{align}
\begin{split}
    H'_{\pm}(\tau) &=V\cos(\omega \tau)\,T_{\pm}\\
    T_{\pm} &\coloneq \sum\limits_{i}a_{i}^{\dagger}a_{i+\hat{x}} \pm a_{i}^{\dagger}a_{i+\hat{y}}+\text{H.C.},    
\end{split}
\end{align}
\noindent where $\tau$ is the evolution time, and $V$ and $\omega$ are, respectively, the amplitude and frequency of the modulation.
$T_\pm$ collects terms involving nearest-neighbor tunneling, with $\pm$ indicating a relative sign between the $x$ and $y$ tunneling terms.
Modulation involving $T_+$ corresponds to the $A_1$ Raman response, and $T_-$ to the $B_1$ response.
We define susceptibilities associated with $T_\pm$:
\begin{equation}
    \chi_{\pm}(\omega) = -i\int_{-\infty}^{\infty} d\tau\, \theta(\tau) \braket{[T_{\pm}(\tau),T_{\pm}(0)]} e^{-i\omega \tau},
\end{equation}
 where $\braket{\cdots}$ indicates a thermal expectation value.

Under linear response, the rate of energy increase due to modulation by $H'_\pm$ is related to the imaginary part of $\chi_\pm$:
\begin{equation}
\label{eq:heating}
    \left. \overline{\frac{dE}{d\tau}}\right\rvert_{\omega, \pm} = \frac{1}{2} \omega |V|^2 \Im \left(\chi_{\pm}(\omega)\right),
\end{equation}
where the overline indicates averaging over a modulation period.
Note that to remove the trivial frequency dependence in Eq.~\eqref{eq:heating}, we scale $V$ by $1/\sqrt{\omega}$ throughout this work, such that a change in the observed heating rate can be associated with a change in $\Im\left(\chi_\pm(\omega)\right)$.

After the above modulation and associated heating, we perform a quasiadiabatic ramp back to a band insulator (BI) and measure the number of defects (singly occupied sites) in the BI, which provides an estimate of entropy.
For equal initial temperatures, and for sufficiently small amounts of heating, the increase in the number of defects in the BI is proportional to $\Im\left(\chi_\pm(\omega)\right)$, and thus proportional to an electronic Raman response in a solid state context.
We refer to this increase in defect number as `counts.'
In situations where we vary the modulation strength $\frac{1}{2}\omega(V/t)^2\tau$, we normalize this number to a condition where $\frac{1}{2}\omega|V/t|^2\tau=0.073$, and refer to this as `normalized counts.'

\placefigure[!ht]{SI-linearity}

To check the validity of Eq.~\eqref{eq:heating}, we measure $\Im(\chi_{\pm}(\omega))$ versus (dimensionless) modulation strength $\frac{1}{2}\omega (V/t)^2 \tau$ for several different drive durations $\tau$, and at a frequency of $\omega = 2\pi\times1200\rm{Hz}$ (Fig.~\ref{fig:linearity}).
We perform this test on a sample close to half-filling, in the pseudogapped region identified in Fig.~\ref{fig:U8_raman}, where $\omega=2\pi\times 1200\rm{Hz}$ is close to the location of the strongest response (i.e. the two-magnon peak).
We observe responses in drive strength which are approximately linear and, up to changes in the background due to varying hold time, similar for different drive durations.

When measuring systems at high temperatures, it is natural to expect a decrease in the number of defects induced by modulation than in cold systems since, thermodynamically, temperature is the derivative of energy with entropy.
In Fig.~\ref{fig:linearity}c, we plot the defect number versus hold time in the final state under typical spectroscopic conditions.
We observe a weak saturation of heating versus the hold time, which we empirically model by fitting a hyperbolic tangent plus an offset.
In the one spectroscopic dataset we present at higher temperatures (Fig.~\ref{fig:u8B1}c), we normalize the raw experimental signal (counts above background) by the indicated slopes of the fitted function in Fig.~\ref{fig:linearity}c.
In doing so we assume that the dominant heating mechanism in Fig.~\ref{fig:linearity}c can be modeled as a constant increase in energy vs time.
We believe the heating is dominated by off-resonant photon scattering, for which this is a reasonable approximation.
Fig.~\ref{fig:linearity}c also gives a sense of the ratio of signal strength to background heating in our typical spectroscopic durations of $30$ms and $60$ms.

\placefigure[!ht]{SI-driven-corr}

\subsubsection{Driven correlations}
\label{subsec:drivencorr}
In addition to measuring a heating response by reverting the system to a BI, we can also use the quantum gas microscope to measure the response of various physical observables to the  lattice modulation.
Specifically, via the Kubo formula~\cite{bruus_flensberg}, the time variation of an observable $\mathcal{O}$ at time $\tau$ after a drive $H'_{\pm}(\tau)=f(\tau)T_{\pm}$ is applied should be given by
\begin{equation*}
    \langle\mathcal{O}\rangle(\tau)-\langle\mathcal{O}\rangle(0)=\int _{-\infty}^{\infty}d\tau' \chi_{\mathcal{O},T_{\pm}}(\tau-\tau')f(\tau')
\end{equation*}
where $\chi_{\mathcal{O},T_{\pm}}(\tau)=-i\theta(\tau)\langle[\mathcal{O}(\tau),T_{\pm}(0)]\rangle$ is the response function of $\mathcal{O}$ to $T_{\pm}$.
For a harmonic drive $f(\tau)=V\cos(\omega \tau)$, at sufficiently long times, $\langle\mathcal{O}\rangle(\tau)$ will also be harmonic, according to $\langle\mathcal{O}\rangle(\tau)=\Re{[\chi_{\mathcal{O},T_{\pm}}(\omega)\exp(-i\omega \tau)]}$, where $\chi_{\mathcal{O},T_{\pm}}(\omega)$ is the Fourier transform of $\chi_{\mathcal{O},T_{\pm}}(\tau)$.
This is strictly true in the infinite-time limit, while for finite drive times $T$, $\chi_{\mathcal{O},T_{\pm}}(\omega)$ really represents an average over a frequency window $\sim 2\pi/T$.
Thus one can measure $\chi_{\mathcal{O},T_{\pm}}(\omega)$ by first measuring $\langle\mathcal{O}\rangle(\tau=0)$, then $\langle\mathcal{O}\rangle(\tau=2\pi N/\omega)$, then $\langle\mathcal{O}\rangle(\tau=(2\pi N+\pi/2)/\omega)$, where $N$ is an integer; the latter two measurements respectively sense $\Re[\chi_{\mathcal{O},T_{\pm}}(\omega)]$ and $\Im[\chi_{\mathcal{O},T_{\pm}}(\omega)]$ as a difference from the first measurement.

In Fig.~\ref{fig:U8_raman}d, we perform the above measurement at a frequency close to the two-magnon peak and measure both spin and density correlations.
That is, in the upper panel of Fig.~\ref{fig:U8_raman}d, $\mathcal{O}$ takes values of $S^z_{i+\hat{x}}S^z_i$, where $S^z_i$ is the $z$-component of the spin of site $i$.
In the middle panel, $\mathcal{O}$ takes values of $S^z_{i+\hat{x}}S^z_i$, where $\vec{d}$ is allowed to vary.
In the lower panel, $\mathcal{O}$ takes values of $m_{i+\vec{d}}m_i$, where $m_i$ is the local moment (singles density).
For the sake of robustness, rather than take three points as described above, for the data in Fig.~\ref{fig:U8_raman}d we measure five values of $\tau$ that span one cycle of the drive, and fit a sine wave $\langle\mathcal{O}\rangle(\tau)=C+A\cos(\omega \tau+\phi)$ with a variable amplitude, offset, and phase to the data (Figs.~\ref{fig:driven-corr},~\ref{fig:U8_raman}d).
The response function can then be inferred as $\chi_{\mathcal{O},T_{\pm}}=Ae^{-i\phi}$.
The data in the lower panels of Fig.~\ref{fig:U8_raman}d shows the magnitudes $A$ of $\chi_{\mathcal{O},T_{\pm}}$.

\subsection{Imaging procedure and fidelities}\label{subsec:imaging}
We take 4 kinds of images in our sequence: (i) parity projected images in sparse lattice (ii) number resolved images in sparse lattice (iii) parity projected images in the dense lattice, and (iv) doublon images in the dense lattice. 

For all of the images, we rapidly freeze the tunnelings by ramping up the lattice depth to $\sim 70 E_r$ and hand off to high power imaging lattices, where we perform fluorescence imaging with Raman sideband cooling. 
When there are two atoms on the same site prior to imaging, the pair of atoms is lost through light-assisted collisions and appears as a hole. In the sparse lattice we can prevent parity projection by maintaining strong repulsion between the atoms while adiabatically ramping up a dense lattice that splits each site into two sites. Since there are at most two atoms on each sparse lattice site, we end up with at most one atom on each dense lattice site, thus preventing parity projection.

However, for atoms loaded in the dense lattice, we cannot prevent parity projection in the same way since we do not have a shorter spacing lattice to further split the sites. 
Thus we do not have access to the number resolved snapshots of the final state. However, we can still measure the average density of each site in the final state by combining two ensembles of images - (i) with the usual parity projection, and (ii) with doublon readout.
For doublon readout, we use a protocol similar to Ref. \cite{mitra_attractive_hubbard}. After freezing the tunneling we transfer the singly occupied sites with state 2 to state 3 via an interaction-resolved RF Landau-Zener (keep doublons as 1-2 doublons, transfer singles 2$\rightarrow$3), and blowout states 1 and 3 via a resonant spin removal beam. This removes all the singly occupied sites and leaves behind the state 2 atoms from the doublons with a fidelity of $79.1(4)\%$ (characterized by performing the same procedure on a band insulator). We calibrate the doublon readout fidelity interleaved with the actual measurements ($\sim 2.6\%$ of experimental shots are doublon calibration) to track drifts of the fidelity over time. The fidelity fluctuates about $\pm2\%$ from day-to-day.

The imaging fidelity comprises of the binarization fidelity and the imaging survival probability. We measure the imaging survival by taking two standard images of a sparsely filled lattice with a variable number of additional imaging pulses in between the two images, and fit a slope and offset to the hopped fraction and lost fraction \cite{parsons_imaging}. We obtain an average imaging survival probability of $F=0.989(1)$.

\subsection{Calibration of experimental parameters}\label{subsec:calibration}
\subsubsection{Calibration of $t$ and $U$}
\label{subsubsec:bandgap_spectroscopy}

In past work, we have calibrated the Hubbard model parameters $t$ and $U$ through a comparison to quantum Monte Carlo data~\cite{cooling_paper,muqing_triangle}.
Recently, discrepancies~\cite{cooling_paper} between experimental data and CP-AFQMC data at low temperatures have highlighted the importance of experimental calibration procedures which are independent of many-body methods, as well as a better understanding of experimental systematics, especially when studying regimes that are challenging to simulate numerically.
Therefore in this work, wherever possible, we attempt to use fully \textit{ab-initio} calibrations of the experimental parameters.

We calibrate the values of tunneling and interaction integrals in the lattice through lattice modulation spectroscopy, as in e.g. Ref.~\cite{daniel_mott}.
In brief, we modulate the intensity of one or both of the lattice beams, which produces interband excitations when the modulation frequency is resonant with a bandgap.
These excitations can be measured by counting the number of atoms that escape the central region of the trap where they are prepared as described in Section~\ref{subsec:preparation}.
We measure many such resonances as a function of the two lattice powers, and fit each one with a Lorentzian to determine a resonance frequency.
We note that the use of a Lorentzian is purely empirical: the lineshape should in principle depend on the bandstructure, but all transitions are measured in a (deep-lattice) regime where the bandwidths are so small that we believe other effects (e.g. alignment drifts between repetitions, power broadening) dominate the linewidth.

\placefigure[!ht]{SI-bandgap}

Through a bandstructure calculation, we fit a model of the lattice potential to these resonances.
Specifically, for each lattice beam, we fit the $3d$ angle of the beam relative to the atoms, as well as a linear conversion factor that relates power, measured in Volts on a photodiode, to potential depth, in Joules on the atoms. Fig.~\ref{fig:bandgap} shows the resonance frequencies measured (i.e. the centers of the Lorentzian fits, upper panel) together with the residuals from the bandstructure fit (lower panel).

Using these fitted values, we calculate the bandstructure of the lattice in the condition $(V_{\bar{X}}=0.557\rm{V}$, $V_Y=0.500\rm{V})$
used for the compressibility and lattice modulation spectroscopy measurements.
We calculate the Hubbard parameters from this information following Ref~\cite{daniel_thesis}.
Specifically, we calculate tunneling energies from a Fourier transform of the bandstructure.
We define $t=(t_{1,0}+t_{0,1})/2=1112(4)\rm{Hz}$ as the arithmetic mean of the nearest neighbor tunnelings, and use this as the unit of energy throughout the measurements.
We obtain Wannier functions for the lattice as described in Ref.~\cite{daniel_thesis}, and calculate the Hubbard interaction $U$ through an overlap integral between these Wannier functions and a contact potential with a strength set by the $s$-wave scattering length $a_s$~\cite{daniel_thesis}.
We infer $a_s$ from the magnetic bias field we apply and the high-precision scattering length data in Ref.~\cite{zurn_scattering}.
We calibrate our magnetic bias field by performing rf spectroscopy on the atomic hyperfine spin.
For reference, we tabulate the bias fields and corresponding interaction strengths, labeled $(U/t)_{\rm{Wannier}}$, used in this work in Table~\ref{tab:interactions}.

\begin{table}
\centering
\begin{centering}
\begin{tabular}{ | c | c | c |}
\hline
Bias field & $(U/t)_{\rm{Wannier}}$ & $(U/t)_{\rm{QMC}}$ \\
\hline
554.64 G & \Ucal & \Uqmccal \\
556.64 G & \Uiii & \Uqmciii \\
564.64 G & \Uiv & \Uqmciv \\ 
572.64 G & \Uv & \Uqmcv \\
580.64 G & \Uvi & \Uqmcvi \\
589.64 G & \Uviii & \Uqmcviii \\
609.64 G & \Uxi & \Uqmcxi \\
\hline
\end{tabular}
\end{centering}
\caption{\textbf{Experimental bias fields and calculated interaction strengths.}}
\label{tab:interactions}
\end{table}

We note that in the bandstructure calculation used above, it is important to use a fully $3d$ model of the lattice potential that accounts for the non-separability of the potential in all three spatial directions.
Prior work (e.g. Ref.~\cite{daniel_mott}) has generally approximated the lattice as being separable in the in-plane and out-of-plane directions.
We find, however, that accounting for the full $3d$ non-separability gives a substantially better fit to the spectroscopic data, and can affect the calculated Hubbard parameters at a $\sim 10\%$ level.

The fitting procedure described above that matches the lattice potential model to spectroscopic data is a least-squares fit.
In this fit, we use the linewidth of each resonance as an errorbar, rather than the errorbar on the fitted center inferred from the Lorentzian fit.
This is a conservative estimate of the uncertainty on each resonance frequency, since the errorbars on the fitted centers are much less than a linewidth.
However, we believe we operate in a regime where systematic uncertainties (e.g. alignment drifts) dominate statistical uncertainties, and these are better captured by the linewidth.
These uncertainties on the resonance frequencies are used to generate uncertainties on the fitted parameters (angles and power conversions) of the lattice through the least-squares fit.
We report uncertainties on the Hubbard parameters by linearly propagating the uncertainties on these fitted parameters through the bandstructure calculation, using finite difference.
As is evident from the fit residuals in Fig.~\ref{fig:bandgap}, this is likely a conservative estimate of the error.

As described in prior work~\cite{cooling_paper}, the lattice beams reflect off a superpolished substrate $\sim 10\mu\rm{m}$ above the atoms.
We define the `vertical' direction to be normal to this substrate.
By symmetry, one expects that the bandstructure fit should not be able to resolve global rotations of both beams about the vertical axis.
The quality of our fit in fact weakly depends on this global azimuthal angle.
We believe this is because in our model we assume the polarization of the lattice light is perfectly linear and polarized in a plane parallel to the vertical axis.
The polarization of the light affects the lattice potential both due to the finite interference angle of the beams on the substrate (a change in polarization changes interference contrast), and due to polarization-dependent losses due to Fresnel reflection on the glass cell that holds the atoms in vacuum.
The sensitivity of the fit to the global angle, however, is rather weak.
For the calculation of Hubbard parameters, we assume that we operate in a condition where the azimuthal angle between $\bar{X}$ and the glass cell is $45\rm{deg}$.
We also assume that the $\bar{X}$ and $Y$ lattice have the same vertical angle relative to the substrate, since deviations from this condition cause the combined potential to be aperiodic in the vertical direction, complicating the calculation.
Individually fitting spectroscopic data from the $\bar{X}$ and $Y$ lattices indicate these angles are matched to within $0.1(1)$ degrees.
With these constraints, the model fitted in Fig.~\ref{fig:bandgap} has 4 free parameters.

\begin{table}
\centering
\begin{centering}
\begin{tabular}{ | c | c |}
\hline
Term & Value \\
\hline
$U_{0001}$ & -0.0616(3) \\
$t_{1,0}$ & 1.021(4)\\
$t_{0,1}$ & 0.979(4)\\
$t_{2,0}$ & -0.0329(3)\\
$t_{0,2}$ & -0.0300(3)\\
$t_{1,1}$ & 0.006(2)\\
$t_{0,3}$ & 0.00148(2)\\
$t_{3,0}$ & 0.00170(3)\\
$t_{2,1}$ & -0.0004(1)\\
$t_{1,2}$ & -0.0003(1)\\
\hline
\end{tabular}
\end{centering}
\caption{\textbf{Deviations of experiment from `vanilla' Hubbard model.} Parameters are listed in units of the tunneling energy $t$.
$U_{0001}$ is calculated assuming a bias field of $589.64$ G.}
\label{tab:deviations}
\end{table}

From the bandstructure calculation, we can estimate several deviations of the experiment from the `vanilla' Hubbard model with only nearest-neighbor tunneling and an on-site interaction, which we list in Table~\ref{tab:deviations}.
The $x$ and $y$ tunnelings are not quite equal.
There are also small but finite longer-range tunnelings.
The contact interaction responsible for the Hubbard interaction can also produce small off-site interaction terms.
The largest such terms involve three Wannier functions on one site and one on a neighboring site, and thus are sometimes given the name `density-assisted tunneling'~\cite{white_density_assisted}.
We report the value of the largest such term, which we denote $U_{0001}$, under conditions where $U/t=\Uviii$.
By this calculation, the leading deviations from `vanilla' Hubbard behavior are the second-neighbor tunnelings $t_{2,0},t_{0,2}\sim-0.03t$ and $U_{0001}\sim-0.06t$.

We note that such deviations are potentially important when comparing our data to theories of the weak-coupling pseudogap~\cite{vilk_tremblay_1997,schafer_footprints}.
For instance, the difference between $x$ and $y$ tunneling integrals breaks the perfect nesting of the square lattice at half-filling, which might weaken the antiferromagnetic fluctuations that generate the pseudogap.
It is also known that the `Vilk criterion' for the onset of the weak-coupling pseudogap is modified in the presence of a Van Hove singularity~\cite{vilk_tremblay_1997}.
Changes to the bandstructure that affect the Van Hove singularities may thus be relevant.
Deviations of this type may also contribute to the discrepancy found with CP-AFQMC in Ref.~\cite{cooling_paper}.

Recent theoretical work~\cite{parish_levinsen_hubbard} has found that the above computation of the Hubbard $U$, in terms of a Wannier overlap integral and a contact interaction, may be overly simplistic.
In particular, the authors of Ref.~\cite{parish_levinsen_hubbard} find that the value of the Hubbard $U$ needed to reproduce the two-body scattering amplitude of an exact model deviates downwards from the formula typically used in the literature, and which we use here.
This deviation is controlled by the ratio $a_s/l$ of the scattering length to the harmonic oscillator length of a single lattice site, which in our experiment is $\sim 0.1$.
Some of this effect might be captured by the inclusion of off-site interaction terms like $U_{0001}$, though it seems unlikely that this can capture e.g. virtual population of higher bands during collisions.
We have not attempted to implement the calculation of Ref.~\cite{parish_levinsen_hubbard} for our experimental setup due to the mathematical complexity of adapting the calculation to our lattice geometry (in particular, the above-mentioned $3d$ non-separability).
Doing so in future work may refine the precision with which we calibrate $U/t$.

We note also that recent experimental work~\cite{bloch_gates,petar_conversation} has found discrepancies between a procedure similar to the above and an \textit{in-situ} procedure using dynamics in small, well-controlled Hubbard systems.
The proposed cause for this discrepancy is differential thermal effects between the conditions where bandgap spectroscopy is done and where data is taken.
It was not possible in the current work to perform such \textit{in-situ} calibration of $t$ and $U$ due to the lack of adequate controls.
Therefore, as a cross-check on the spectroscopic calibration of $U/t$, within DSLat we also repeated the calibration procedure used in Ref.~\cite{cooling_paper}, namely a comparison of the singles density at half-filling to QMC.
Using this method, at a bias field of $589.64$ G, we measure a maximum singles density of $0.885(3)$, leading us to infer $U/t=7.7(1)$ by comparison with CP-AFQMC data at $T=0$.
We use this value to generate estimates for $U/t$ at the other magnetic bias fields using the assumption that $U/t$ scales linearly with the scattering length.
We tabulate these values in Table~\ref{tab:interactions}, as $(U/t)_{\rm{QMC}}$.
Unless noted explicitly, we use the \textit{ab initio} value $(U/t)_{\rm{Wannier}}$ when quoting values for $U/t$.

\subsubsection{Calibration of $T$}
\label{subsubsec:temperature}
\begin{table*}
\centering
\begin{centering}
\begin{tabular}{ | c | c | c | c | c |}
\hline
Dataset & $n,(U/t)_{\rm{Wannier}}$ & $n_s,(U/t)_{\rm{Wannier}}$ & $n,(U/t)_{\rm{QMC}}$ & $n_s,(U/t)_{\rm{QMC}}$ \\
\hline
Fig.~\ref{fig:widom}b, cold & $0.12(1)$ & $0.14(1)$ & $0.15(2)$ & $0.09(2)$ \\
Fig.~\ref{fig:widom}b, hot & $0.36(1)$ & $0.35(1)$ & $0.365(8)$ & $0.311(8)$ \\
Fig.~\ref{fig:u8B1}c &$0.356(7)$ & $0.342(6)$ & $0.356(5)$ & $0.303(6)$\\
DS0, avg. & $0.146(7)$ & $0.173(5)$ & $0.163(5)$ & $0.12(1)$ \\
DS1, avg. & $0.10(1)$ & $0.169(3)$ & $0.145(4)$ & $0.095(3)$ \\
DS2, avg. & $0.118(4)$ & $0.157(3)$ & $0.140(3)$ & $0.073(5)$ \\
DS3, avg. & $0.11(2)$ & $0.14(1)$ & $0.14(1)$ & $0.08(2)$ \\
\hline
\end{tabular}
\end{centering}
\caption{\textbf{Thermometry.} Different columns show the results of different comparisons against QMC data, described in Section~\ref{subsubsec:temperature}.}
\label{tab:temperatures}
\end{table*}
As in prior work~\cite{cooling_paper}, we calibrate the temperature by comparison to quantum Monte Carlo simulations.
At high temperatures, it is feasible to compare against numerically exact DQMC simulations.
At low temperatures, where DQMC is affected by the fermion sign problem, we compare against state-of-the-art, but approximate constrained-path auxiliary field quantum Monte Carlo (CP-AFQMC) simulations.
Specifically, we compare against the CP-AFQMC and DQMC data used in Ref.~\cite{cooling_paper}, plus additional DQMC data we have produced for this work, mostly at $U/t\leq 7$.

In these comparisons, we rely on the temperature dependence of spin correlations.
Specifically, within QMC, one can compute the singles density $n_s$ and the nearest-neighbor spin correlation $C_{1,0}$ as a function of temperature at the experimental $U/t$.
One can invert this data to obtain a function $T(n_s,C_{1,0},U)$ which is applied to the experimental data.
This procedure was used in Ref.~\cite{cooling_paper} to calibrate the temperatures reached by an earlier version of the cooling scheme used in this work (Section~\ref{subsec:preparation}).

In the current work, we attempt to refine this procedure by using the \textit{ab initio} value of $U/t$ (Section~\ref{subsec:calibration}), and by using the knowledge of the full density (rather than singles density) enabled by the doublon-resolved imaging technique described in Section~\ref{subsec:imaging}.
Access to the former makes the calibration of $U/t$ independent of many-body methods, and improves the statistical errorbar.
Access to the latter allows comparison against different interpolation schemes in the comparison against QMC, specifically the function $T(n,C_{1,0},U)$ (in addition to $T(n_s,C_{1,0},U)$).
The latter is a particularly valuable cross-check, as the singles density at fixed density depends strongly on $U/t$, due to the interaction-dependence of the double occupancy.

In Table~\ref{tab:temperatures}, we show a comparison between different thermometry methods.
Specifically, we compare singles density-based thermometry with density-based thermometry, and also compare results for the spectroscopic $U/t$ calibration with the QMC $U/t$ calibration (see Section~\ref{subsec:calibration}).
Evidently the different methods give results which are similar, but not identical.
The trends versus $U/t$ and $n$ vs $n_s$ interpolation are due to the fact that at fixed density (in our range of $U/t$), increasing $U/t$ increases the nearest-neighbor correlation, while at fixed singles density, increasing $U/t$ decreases nearest-neighbor correlation (because it amounts to decreasing density).
In Fig.~\ref{fig:widom} and the surrounding discussion we report the maximum of the four `cold' numbers for the `cold' temperature, to give a conservative estimate of temperature.
For the `hot' temperature we average two $(U/t)_{\rm{Wannier}}$ results (since they are the \textit{ab initio} ones), and include the difference between the two as a systematic uncertainty in the errorbar.
We also in Table~\ref{tab:temperatures} show averaged temperatures for low-temperature data in datasets DS0,DS1,DS2, and DS3, as well as temperatures for the hot data in Fig.~\ref{fig:u8B1}c (which is from DS1).

\subsubsection{Systematic trends in $T/t$ versus $U/t$}\label{subsubsec:T_vs_U}
As discussed in Section~\ref{subsec:collection}, all the data have interleaved spin correlation data as a check that the system remains cold.
This check is performed at a reference condition of $\delta\approx 0.1$ and $U/t=\Uviii$, similar to the conditions studied in Ref.~\cite{cooling_paper}.
However, since the data in this work involves changes in interaction strength and doping, it is valuable to check for systematic trends in temperature as these parameters are varied.
Ref.~\cite{cooling_paper} already found little trend in temperature versus doping, but trends versus interactions have not been checked.
To this end, we measured spin correlation data versus $U/t$ at a roughly constant density $\delta\approx 0.025-0.05$ in a dedicated dataset DSTherm (see Section~\ref{subsec:collection}).
We analyze the temperatures obtained using the four methods described above, and plot the results in Fig.~\ref{fig:T_vs_U}.

\placefigure[!ht]{SI-T-vs-U}

As in Table~\ref{tab:temperatures}, the inferred temperature depends rather strongly on the method of thermometry.
We note in this connection that the systematics described in Section~\ref{subsubsec:temperature} become even more severe as the interaction is lowered: the temperature-dependence of the spin correlations becomes reduced, and the interaction-dependence of the singles density at fixed density grows.
The large errorbars at low $U/t$ in the QMC-based methods, in particular, are dominated by the statistical error on $U/t$ in this calibration method.

Thus, while the data upper bound any changes in temperature caused by changing $U/t$, it is challenging to make quantitative statements about trends, due to the difficulty of thermometry.
In such an upper bound, it is important to note that the `reference' condition at $U/t=\Uviii$ in DSTherm happens to be comparatively hot, as is evident from comparing Fig.~\ref{fig:T_vs_U} with Table~\ref{tab:temperatures}.

Understanding potential trends in temperature versus $U/t$ is important for using the experimental data to establish differences between the strong- and weak-coupling pseudogap, of the type suggested in the main text.
If all the experimental data is as cold as, for instance, the data in Ref.~\cite{cooling_paper}, then even at the lowest values of $U/t$ studied in this work, we should encounter the regime found by DiagMC where the one-particle function has a pseudogap~\cite{simkovic_pseudogap}.
In this case, the absence of e.g. a compressibility peak in our data would imply a difference between the weak- and strong-coupling pseudogap.
However, if the experimental data at low $U/t$ is substantially hotter than at large $U/t$, it might exit the pseudogapped regime found in DiagMC, nullifying the comparison.

In addition to improved thermometry, one way to make progress on this issue is direct experimental measurement of the one-particle function~\cite{waseem_arpes}, so as to remove the need for a comparison with DiagMC.
Alternatively, future numerical work could study the compressibility and two-particle response functions at low temperatures in the range of low $U/t$ where DiagMC finds a pseudogap, but where our experiment does not.
We note in this context that lower values of $U/t$ are often easier to study numerically than large values, depending on the computational scheme~\cite{qin_computational_review}.

\subsection{Data analysis}\label{subsec:analysis}
\subsubsection{Densities and correlation functions}
All the density measurements and correlation functions reported in this work are averaged over uniform regions of the potential described in Sections~\ref{subsec:preparation} and~\ref{subsec:compressibility}.
Since densities are a direct mean of snapshot data, errorbars on density are computed as the standard deviation of the mean.
Errorbars on spin correlations are computed via bootstrap, as described in previous work~\cite{cooling_paper}.
The lattice modulation spectroscopy data is obtained by taking parity projected images (see Section~\ref{subsec:imaging}) of the final band insulator and averaging the number of singly occupied sites.
This is also a direct mean of snapshot data, and errorbars are computed as the standard deviation of the mean.

\subsubsection{Converting singles density to density}
\label{subsec:conversion}
As mentioned in Section~\ref{subsec:collection}, datasets DS0 and DS1 do not use the doublon-resolved readout described in Section~\ref{subsec:imaging}, but only parity-projected images.
Reporting densities in these two datasets thus requires knowledge of the function $n(n_s)$.
On the basis of DQMC simulations at elevated temperature (Fig.~\ref{fig:combination_fig}b), as well as CP-AFQMC simulations at low temperatures~\cite{cooling_paper}, we expect $n(n_s)$ to be a comparatively smooth function with little temperature dependence.
The $n(n_s)$ function, however, has substantial interaction dependence, due to the interaction dependence of the double occupancy.
We therefore measure the $n(n_s)$ function experimentally at all the interaction strengths studied in this work in a dedicated dataset, DSConv, using a combination of doublon-resolved and parity-projected images.
We also use the data in DS3 for this purpose, since it is a large collection of doublon-resolved and parity-projected snapshots versus density and interactions.
To obtain the $n-n_s$ conversion, we fit a quadratic function to the data using orthogonal distance regression.
This function is used to convert $n_s$ measurements in DS1 to $n$ measurements (e.g. the doping axes of Fig.~\ref{fig:U8_raman}).
We plot the resulting quadratic fits in Fig.~\ref{fig:combination_fig}d (gray lines in the third row from the top).

Like the spectroscopic dataset DS1, the compressibility dataset DS0 only uses parity-projected images, and thus can only access the singles density $n_s$.
DS0 thus really measures the singles compressibility $\kappa_s=dn_s/d\mu$.
In principle, knowledge of the $n(n_s)$ function allows conversion of this quantity to $\kappa=dn/d\mu$ via the derivative $dn/dn_s$.
However, because it involves a derivative, the numerical details of extracting this conversion are more sensitive than in the $n(n_s)$ conversion used in DS1.

In particular, the functional form of the $n(n_s)$ function is not known \textit{a priori}.
However, particle-hole symmetry in the Hubbard model dictates that $n(n_s)$ is symmetric across $n=1$, causing a divergence in $dn/dn_s$.
DQMC simulations at elevated temperatures (Fig.~\ref{fig:combination_fig}b,c) indicate that this divergence onsets at larger dopings as the interaction strength is decreased.
This is compatible with the fact that at $U=0$, $n_s=n(1-n/2)$ due to the absence of correlations between the two spin states, while at large $U$, $n\approx n_s$ due to the suppression of double occupancies (Fig.~\ref{fig:combination_fig}b).
It is challenging to obtain experimental data of sufficient quality to capture the $dn/dn_s$ conversion without prior assumptions on the functional form of $n(n_s)$.
Restricting the functional form of $n(n_s)$ leads to systematic errors, which cause large errors in $dn/dn_s$ in the regime where $dn/dn_s$ is divergent.
Conversely, we find that using a high-order interpolant on the experimental data to more accurately capture behavior close to half-filling leads to overfitting at large doping, causing significant oscillations in the inferred $dn/dn_s$ that depend on the interpolation scheme.
Similarly, despite the mild temperature dependence of $n(n_s)$, it is not practical to use high-temperature DQMC data to compute $dn/dn_s$, because details of the divergence depend sensitively on the value of $U/t$, which is only known with finite precision (see Section~\ref{subsec:calibration}).

Our approach is thus to only estimate $dn/dn_s$ at comparatively large dopings, where it is not divergent, and to discard data from DS0 which is at a lower doping.
This amounts to discarding data from DS0 in the regime where the measurement loses sensitivity due to the divergence in $dn/dn_s$.
Restricting the fit to large doping allows the use of a more restrictive fit  (see Section~\ref{subsubsec:combination}), due to the comparatively smooth behavior of $n(n_s)$ at large doping.

\subsubsection{Combining DS0 and DS3}\label{subsubsec:combination}

Since datasets DS0 and DS3 have comparable statistical power, it is desirable to average them together to report a final value for the compressibility.
Motivated by the considerations in Section~\ref{subsec:conversion}, we do so using the following procedure.
At each experimental interaction strength, we compute $n(n_s)$ using DQMC at high temperatures.
We use these simulations to extract a value $n_s^{\rm{cut}}$ at which a linear approximation to $dn/dn_s$ vs $n_s$ for $n_s<n_s^{\rm{cut}}$ incurs a $5\%$ relative error.
We use orthogonal distance regression to fit the experimental $n(n_s)$ data from DSConv and DS3 to a quadratic for the data satisfying $n_s<n_s^{\rm{cut}}$ (Fig.~\ref{fig:combination_fig}).
We numerically differentiate this quadratic to obtain $dn/dn_s$.
This analysis is illustrated in Fig.~\ref{fig:combination_fig}c for all interaction strengths studied.
Having found $dn/dn_s$, we sample this function at the $n_s$ values measured in DS0, and multiply each $\kappa_s$ measured in DS0 by its corresponding $dn/dn_s$ value to obtain $\kappa$ from DS0.

\placefigure[!ht]{SI-kappa-combination}

This produces the data shown in Fig.~\ref{fig:combination_fig}d.
To average the datasets together, we average together points lying in doping bins of width $0.04$.
This value is chosen to be larger than the typical density difference imprinted by the step $\Delta\mu$, as well as the typical variation of the density in the homogeneous region due to disorder (see Section~\ref{subsec:preparation}).

When estimating errors in this analysis, it is important to account for statistical uncertainty in the calibration of $\Delta\mu$ for points from DS0, but not from DS3, because the $\Delta\mu$ in DS3 can be calibrated using many more samples (see Section~\ref{subsec:compressibility}).
Since the same $\Delta\mu$ is used to normalize all points of a single $\kappa(n)$ curve from DS0, when averaging in a bin we first average all $\Delta n$ from DS0 together, then normalize by $\Delta\mu$, then average this with the $\kappa$ from DS3 in the same bin, propagating all errors linearly.
This order of operations reflects the fact that different $\Delta n$ data points from DS0 are statistically independent, but different $\kappa$ data points are not, due to the $\Delta\mu$ calibration.

To maximize the statistical power of the average, we weight the average between DS0 and DS3 by the number of repetitions $N_0$ and $N_3$ that produced each measurement, i.e. we report $(N_0\kappa_0+N_3\kappa_3)/(N_0+N_3)$.
In $N_3$ we only include parity-projected images to reflect the fact that the density of samples is effectively halved in DS3 compared to DS0 due to the addition of doublon-resolved images.
We reduce $N_0$ to account for the uncertainty in $\Delta\mu$ according to $N_0=(1/N_{\Delta\mu}+1/N_{\Delta n})^{-1}$.
We also reduce $N_0$ to account for the sensitivity loss due to $dn/dn_s>1$, according to $N_0\to N_0/(dn/dn_s)^2$. This procedure is intended to make the weights $N_{0,3}/(N_0+N_3)$ reflect an \textit{a priori} estimate of the inverse square of the uncertainties of $\kappa_{0,3}$, which should minimize the final uncertainty.

We note that the values of the compressibility maximum and the likelihood ratio test $p$-values (Section~\ref{subsubsec:widom_fit}) do not depend significantly on the $5\%$ relative error tolerance used above.
In fact, similar results are obtained by performing the fit on DS3 alone.
However, given the relative size of the datasets (Section~\ref{subsec:collection}), we believe it is valuable to use DS0 to the extent possible to maximize the statistical power of the data.

\subsubsection{Fitting the compressibility maximum}\label{subsubsec:widom_fit}

\placefigure[!ht]{SI-widom-fit}

To quantitatively analyze the compressibility data shown in Fig.~\ref{fig:widom} and Fig.~\ref{fig:combination_fig}, we fit the data to several models.
One simple choice is to model the compressibility as a high-temperature `background' compressibility, plus a Gaussian capturing the compressibility maximum:
\begin{align}\label{eq:widom_dqmc_fit}
    \kappa(n)=\kappa_{\rm{DQMC}}(n)+\kappa_2e^{-(n-n_1)^2/\Delta n^2}
\end{align}
where $\kappa_{\rm{DQMC}}$ is computed at a temperature easily accessible to DQMC.
We find that $T/t=0.5$ is roughly the lowest temperature at which we can easily simulate all interactions studied in this work (see Section~\ref{subsubsec:dqmc}), limited by $U/t=\Uxi$, and so we use the $T/t=0.5$ compressibility for $\kappa_{\rm{DQMC}}$.
One can perform a likelihood ratio test using this model to quantitatively assess the presence or absence of a peak.
Specifically, one fits the model~\eqref{eq:widom_dqmc_fit} to the data, and compares the likelihood of this fit to the likelihood of a model with no Gaussian (i.e. restricting $\kappa_2=0$).
The fit is a least-squares fit, i.e. we take the log-likelihood of a given datapoint $(n_{\rm{exp}},\kappa_{\rm{exp}})$ with errorbar $\sigma_{\rm{exp}}$ to be $-\frac{1}{2}(\kappa(n_{\rm{exp}})-\kappa_{\rm{exp}})^2/\sigma_{\rm{exp}}^2$.
Due to the uncertainty in $\Delta\mu$ in DS0, the errorbars of different density points are correlated, so when calculating the log-likelihood of all points, we actually use $-\frac{1}{2}(\kappa(n_{\rm{exp}})-\kappa_{\rm{exp}})^T\Sigma^{-1}(\kappa(n_{\rm{exp}})-\kappa_{\rm{exp}})$, where $(\kappa(n_{\rm{exp}})-\kappa_{\rm{exp}})$ is understood as a vector with length equal to the number of data points and $\Sigma$ is the covariance matrix of the data.
Having obtained both fits, we calculate the Wilks statistic as twice the difference of the log-likelihoods of the fits, and compare this statistic to a $\chi^2$ distribution with $3$ degrees of freedom to obtain a $p$-value for the null hypothesis.
We perform all fits without subjecting the data to the binning procedure described in Section~\ref{subsubsec:combination} which produces the points plotted in Fig.~\ref{fig:widom}.
However, we find that the results do not change substantially if performed on the binned data.

In Fig.~\ref{fig:widom_fit}a (top row), we show the result of the fit, plus the null hypothesis $\kappa_2=0$.
In Fig.~\ref{fig:widom_fit}b, we show the null hypothesis $p$-values versus the interaction strength.
We also show the fitted value of the doping for the compressibility maximum.
Errors on this doping are obtained by linear error propagation.

The $p$-values reported by this test are rather small (Fig.~\ref{fig:widom_fit}b, lower panel, inset).
Inspecting the fits, there are perhaps some values of $U/t$ where such confidence in the presence of a peak is justified, but at other values of $U/t$ (especially $U/t=\Uiv$) it is apparent the likelihood ratio test is not so much testing the presence of a local maximum as it is a systematic difference between the experiment and DQMC, which can be partially remedied with the additional Gaussian term.
There could be several reasons for such a discrepancy.
For example, since compressibility tends to increase with decreasing temperature, it is possible that using DQMC data at $T/t<0.5$ would bring DQMC and experiment closer at $U/t=\Uiv,\Uv$.
Alternatively, systematic errors in the experimental compressibility measurement (see Section~\ref{subsec:compressibility}) might make the experimental data deviate from DQMC.
These considerations motivate the use of a different model,
\begin{align}\label{eq:widom_scaled_dqmc_fit}
    \kappa(n)=\alpha\kappa_{\rm{DQMC}}(n)+\kappa_2e^{-(n-n_1)^2/\Delta n^2}
\end{align}
where $\alpha$ is also an adjustable parameter.
Allowing $\alpha\neq 1$ admits the possibility, for example, of ideal-gas behavior in the compressibility ($\kappa\sim 1/T$) between $T/t=0.5$ and lower temperatures, or of global miscalibrations in the experimental step size $\Delta\mu$.
In the second row of Fig.~\ref{fig:widom_fit}a, we show the results of this fit, and show the fitted compressibility maximum dopings and null hypothesis $p$-value in Fig.~\ref{fig:widom_fit}b.
Note that the flexibility granted to the null hypothesis by the parameter $\alpha$ changes the null hypothesis $p$-value at $U/t=\Uiv$ dramatically, bringing it above $0.05$.

To further test of the presence of a compressibility maximum, we model the compressibility as a sum of a quadratic and a Gaussian,
\begin{align}\label{eq:widom_fit}
\kappa(n)&=\kappa_0+\kappa_1(n-n_0)^2+\kappa_2 e^{-(n-n_1)^2/\Delta n^2},
\end{align}
where $\kappa_{0,1,2},n_{0,1},\Delta n$ are adjustable parameters.
This model choice is empirical.
The quadratic part is intended to capture the smooth behavior exhibited in $\kappa(n)$ by high-temperature DQMC, while placing fewer restrictions on the null hypothesis.
We show the result of this fit in the lower row of Fig.~\ref{fig:widom_fit}a, as well as the $p$-values for the null hypothesis and compressibility maximum dopings in Fig.~\ref{fig:widom_fit}b.
In this test, the $p$-value of the null hypothesis associated with $U/t=\Uiv$ becomes even higher.

The values of the compressibility maximum doping used in the main text and the fit in Fig.~\ref{fig:widom}b are obtained using the model~\eqref{eq:widom_fit} instead of~\eqref{eq:widom_dqmc_fit} or~\eqref{eq:widom_scaled_dqmc_fit}.
Due to the flexibility it grants to the null hypothesis, we believe it is a more stringent test of the power of the experimental data to resolve the presence of a peak.
The comparison with DQMC shown in Fig.~\ref{fig:widom_fit} is still useful, however, in identifying trends versus temperature.
For instance, it is more apparent in the comparison against DQMC than against the quadratic null hypothesis~\eqref{eq:widom_fit} that several points at $U/t=\Uvi$ near $\delta=0.1$ lie systematically above the high-temperature compressibility.

\subsubsection{Fitting the $B_1$ pseudogap}\label{subsec:raman_fit}
To extract the `characteristic' doping associated with loss of response in $B_1$ symmetry in the pseudogap, we fit the $B_1$ response versus doping $\delta$ to a sigmoid functional form
$f(\delta)=C+A/(1+e^{-(\delta-\delta_0)/\Delta\delta})$ where $C,A,\delta_0,$ and $\Delta\delta$ are adjustable parameters.
We then define $\delta_{\star}=\delta_0+2\Delta\delta$ to capture the `onset' point for the loss of response.
This definition of $\delta_{\star}$ locates the `onset' at the intersection of a horizontal asymptote to the large-doping value of the response, and a line tangent to the response at the halfway point.
We show the fits in Fig.~\ref{fig:B1_fit}.

\placefigure[!ht]{SI-B1-fit}

\subsubsection{Fitting the two-magnon peak}
\label{subsec:2mp_fit}
To extract a peak frequency for the two-magnon peak, we fit a quadratic to the data versus frequency, restricting the fitting range to the points above half-maximum. For the lowest interaction strength, the peak shape has significant skew and hence we fit the data with value greater than 0.75 times the peak value. We report the frequency of the maximum of this fit as the peak, and the errorbars are the uncertainty of the fitted peak position.

\begin{table*}
\centering
\begin{centering}
\begin{tabular}{ | c | c | c | c |}
\hline
 $U/t$ & Pseudogap onset doping $\delta_*$ & Half-filling two-magnon peak $\omega_{2\text{mg}}$ & Half-filling $U$ peak $\omega_U$ \\
\hline
 \Uiii & No gap onset & 0.33(3) & No peak\\
 \Uiv & Not measured & 0.71(3) & No peak\\
 \Uv & 0.094(7) & 0.97(5) & 5.6(3)\\
 \Uvi & 0.10(1) & 1.22(3) & 6.6(2)\\
 \Uviii & 0.116(9) & 1.27(3) & 7.4(3)\\
 \Uxi & 0.11(1) & 1.01(6) & 10.5(2)\\
\hline
\end{tabular}
\end{centering}
\caption{\textbf{Summary of spectroscopy results.}}
\label{tab:spectroscopy}
\end{table*}

\subsection{Extended data}

\subsubsection{Spin correlation length in the pseudogap}
As discussed in the main text, the spin correlation length is an important quantity in the physics of the pseudogap.
At weak coupling in the Hubbard model, the presence of a pseudogap is controlled by the `Vilk criterion'~\cite{vilk_tremblay_1997,schafer_footprints} that the spin correlation length exceed the thermal de Broglie wavelength of the fermions.
The puzzle of the strong coupling Hubbard pseudogap, and the hole-doped cuprate pseudogap, is that the spin correlation length is short.
To confirm this situation, we measure the spin correlation length at a range of dopings, from the pseudogap to the normal metal.
We show the results of this measurement in Fig.~\ref{fig:corr-length}.
We define $C_{\vec{d}}^{zz}=4\langle S_{i+\vec{d}}^{z}S_i^z\rangle$ as the spin correlation function at displacement $\vec{d}$, as in prior work~\cite{cooling_paper}.
Fig.~\ref{fig:corr-length}a,b show correlations at $\delta=0.04$, in the pseudogap, where we fit a correlation length of several lattice sites ($\xi\sim 3$).
This is comparable to the value measured in the cuprate $\mathrm{La}_{2-x}\mathrm{Sr}_x$Cu$\mathrm{O}_4$ at or below room temperature~\cite{kastner_cuprate_review}.
In Fig.~\ref{fig:corr-length}, we plot the correlation length versus doping measured in the experiment, as well as in DQMC simulations at high temperature.
While $\xi$ generally grows towards half-filling, it does not exhibit particularly sharp behavior compared to the high-temperature curves.

\placefigure[!ht]{SI-corr-length}

\subsubsection{U-scale lattice modulation}
\placefigure[!ht]{SI-u8B1} Fig.~\ref{fig:u8B1} shows additional lattice modulation data in $B_1$ symmetry, both from the experiment and from ED simulations.
Fig.~\ref{fig:u8B1}a shows experiment data versus doping and frequency at $U/t=\Uviii$.
At high energies, close to half-filling, we see a resonance at an energy $\omega\approx U$ due to doublon-hole production~\cite{tillman_doublon}, which narrows and disperses slightly as the doping is increased before eventually vanishing.
We see good agreement with ED simulations performed at the experimental conditions (Fig.~\ref{fig:u8B1}b).
At lower energies, we see the evolution of the two-magnon feature discussed in the main text.
This behavior is partially reproduced by ED.

In Fig.~\ref{fig:u8B1}d we plot experimental data at half-filling versus $U/t$ and frequency.
This is equivalent to Fig.~\ref{fig:raman_vs_U}c in the main text, but over a wider frequency range.
Fig.~\ref{fig:u8B1}e is the analogous plot from ED.
We observe a two-magnon feature and a doublon-hole feature in both plots.
Fig.~\ref{fig:u8B1}f shows the fitted maximum of the two features as a function of $U/t$, from both ED and experiment.
The doublon-hole feature roughly tracks $\omega=U$, and the non-monotonic dependence of the two-magnon frequency vs $U$ is discussed in the main text.

Evidently there is reasonable agreement between experiment and ED in both comparisons, but with some systematic deviations.
At low $U/t$, the experimental two-magnon peak frequency deviates below the ED prediction.
At large $U/t$, the experimental doublon-hole frequency is slightly below the ED prediction.
There are several potential sources for such deviations.
Finite-size effects in ED, for example, could bias the frequencies of low-energy features like the two-magnon peak at low $U$.
Systematic errors in $U$ and $t$ calibrations could also contribute.
We emphasize this comparison is performed without any fitting, but rather using the \textit{ab initio} values of $t$ and $U$ discussed in Section~\ref{subsubsec:bandgap_spectroscopy}.
The finite experimental temperature and deviations of the experiment from `vanilla' Hubbard behavior (see Section~\ref{subsubsec:bandgap_spectroscopy}) could in principle contribute, though we suspect they are not significant effects.

Fig.~\ref{fig:u8B1}c shows the $B_1$ response versus frequency and doping at $U/t=\Uviii$ and $T/t=0.35(1)$.
It is directly analogous to the $B_1$ data in Fig.~\ref{fig:U8_raman}b, but at high temperatures.
The response is considerably weaker, even accounting for the temperature-dependent normalization discussed in Section~\ref{subsec:spectroscopy}.
The height of the two-magnon peak is decreased.
The pseudogap (loss of low-energy response) is weaker, if at all present.
Both features are restricted to lower values of doping than in Fig.~\ref{fig:U8_raman}b.
This is compatible with the notion that the pseudogap temperature is a decreasing function of doping, as in cuprates and in many studies of the Hubbard pseudogap~\cite{lee_wen_nagaosa_review,norman_pines_kallin_pseudogap,simkovic_pseudogap,qin_computational_review}.

\subsection{Numerical methods}\label{subsec:simulations}

\subsubsection{DQMC}
\label{subsubsec:dqmc}

Determinant quantum Monte Carlo (DQMC) is a numerically exact method that can compute physical properties of the Hubbard model on finite lattices~\cite{qin_computational_review}.
However, under generic conditions, the Monte Carlo is affected by a minus sign problem that increases the statistical error.
The number of Monte Carlo samples needed to reach a fixed accuracy scales with $1/\langle s\rangle^2$, where $\langle s\rangle$ is the average sign.
Asymptotically, at low temperatures and large system sizes, the average sign is expected to scale as $\langle s\rangle \sim e^{-\alpha \beta t L^2}$, where $\beta=1/T$ is the inverse temperature and $\alpha$ is a prefactor that depends on the system being simulated (e.g. $U/t$ and doping)~\cite{troyer_qmc,scalettar_critical,scalettar_geometry}.
Notably, in the square lattice Hubbard model, the sign problem has been found to become particularly severe as the pseudogap is approached, which has been interpreted as evidence for a critical point~\cite{scalettar_critical}.

We perform determinant quantum Monte Carlo (DQMC) simulations using the QUEST package~\cite{QUEST}.
We work mostly on an $8\times 8$ lattice, although for computing the compressibility at $U/t=\Ucal,T/t=0.3$ (see Section~\ref{subsec:compressibility}) and $U/t=\Uviii,T/t=\Twidhot$ (Fig.~\ref{fig:widom}b) we work on $14\times 14$ and $12\times 12$ lattices, respectively, to avoid finite size effects.
Following Ref.~\cite{ctqmc_review}, we use a Trotter time step that varies with temperature and interactions, so that the number of imaginary time slices follows $L=5\beta U$ (up to discretization errors, since $L$ is integer).
We use 5000 warm-up sweeps and 25000 or 50000 measurement sweeps.
In cases where the statistical power is reduced by the average sign, we run between 2 and 100 simulations with different seeds and average them.
We do not operate in conditions where the average sign is $<0.1$.
On these lattice sizes, we find it is challenging to reach converged results below $T/t\sim 0.2$ at $U/t=5$, below $T/t\sim 0.3$ at $U/t=7$, and below $T/t\sim 0.5$ at $U/t=11$.

\subsubsection{Exact Diagonalization (ED)} \label{subsubsec:ED}
As described in sec.~\ref{subsec:spectroscopy}, the $A_1$ and $B_1$ response functions are given by the autocorrelation functions $C(t)=\langle T_{\pm}(t)T_{\pm}(0)\rangle$ where $T_{\pm}$ is the perturbation term which drives nearest neighbor tunneling. We use the Lanczos method \cite{lanczos1950, prelovsek2017} to compute this dynamical correlation function in the ground state on a $4\times4$ cluster with open boundary conditions and periodic boundary conditions. We first obtain the ground state wavefunction $|\Psi_0\rangle$ of the finite cluster using the Lanczos expansion until the ground state energy $\epsilon_0$ is converged (typically $\sim 150$ steps). We then repeat the Lanczos expansion for $M\sim 150$ steps, but starting from the perturbed wavefunction $|\Psi' \rangle=\frac{1}{\alpha} T_{\pm}|\Psi_0\rangle$ where $\alpha$ is a normalization factor. The second Lanczos expansion gives a truncated $M\times M$ representation of the full Hamiltonian $H$  in the Krylov space spanned by the states $\{ |\Psi'\rangle , H|\Psi'\rangle , ... H^{M-1}|\Psi'\rangle \}$. We can then write the autocorrelation function (for $t>0$) as
\begin{equation}
\begin{split}
    C(t)&= \langle O(t) O(0) \rangle_0 \\
    &= \langle\Psi_0 |e^{-iHt}Oe^{iHt}O|\Psi_0\rangle\\
    &=e^{-i\epsilon_0t}\langle\Psi_0 |Oe^{iHt}O|\Psi_0\rangle\\
    &=\alpha^2 e^{-i\epsilon_0t}\langle\Psi' |e^{iHt}|\Psi'\rangle\\
    &= \alpha^2 e^{-i\epsilon_0t} \sum_{n}^M\langle\Psi'|\phi_n\rangle\langle\phi_n|\Psi'\rangle e^{i\tilde{\epsilon}_n t}\\
    &=  \alpha^2 \sum_{n}^M \tilde{v}_{n0}^2 e^{i(\tilde{\epsilon}_n-\epsilon_0) t},
\end{split}
\end{equation}
where $\tilde{\epsilon}_n$ are the eigenvalues corresponding to eigenvectors $|\phi_n\rangle$, and $\tilde{v}_{n0}$ are the eigenvector overlaps obtained by diagonalizing the $M\times M$ representation of $H$. The frequency response function $C(\omega)$ is the Fourier transform of the autocorrelation function $C(t)$, and contains discrete poles at $\omega=\tilde{\epsilon}_n-\epsilon_0$ due to the finite system size in the simulation. To obtain a smooth plot, we broaden these poles with a width of $\eta=0.15t$.

\section*{Data and code availability}
Experimental data and code used in this work
will be made available through Harvard Dataverse.

\begin{dfigure}{SA}
    \noindent
    \includegraphics[width=3.54in]{"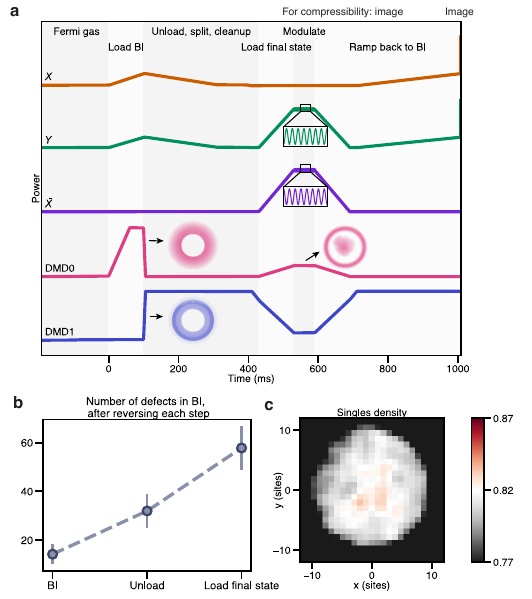"}
    \caption{\textbf{Experimental sequence.} (\textbf{a}) Ramps used in the experimental sequence, described in Section~\ref{subsec:preparation}.
    (\textbf{b}) Number of defects in BI after each stage of the sequence, which gives a semi-quantitative measure of the entropy added in each step (Section~\ref{subsec:preparation}).
    (\textbf{c}) Average singles density in a typical realization of the experiment at $U/t=\Uviii$. A Gaussian blur with $\sigma=1$ is applied to the data to suppress shot noise.    \label{fig:rampfig}}
\end{dfigure}

\begin{dfigure*}{SI-corr-length}
    \noindent
    \includegraphics[width=4.78in]{"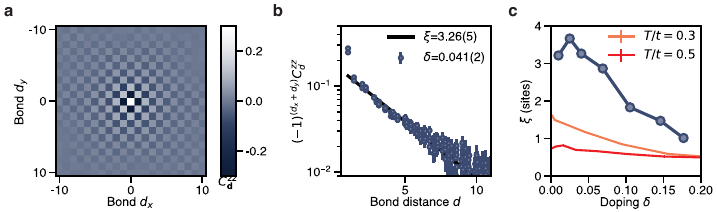"}
    \caption{\textbf{Spin correlation length.} (a) and (b) Spin correlations $C^{zz}_d$ vs bond displacement $d$ at doping $\delta=0.04$. The fitted exponential correlation length $\xi$ is about 3 sites. 
    (c) Correlation length versus doping fit to experimental data (markers) as well as high temperature DQMC data (lines). The experimental correlation length in the pseudogap regime ($\delta < 0.1$) is larger than a lattice site but not comparable to the system size.
    \label{fig:corr-length}}
\end{dfigure*}

\begin{dfigure*}{SI-kappa-combination}
\centering
    \noindent
    \includegraphics[width=7.08in]{"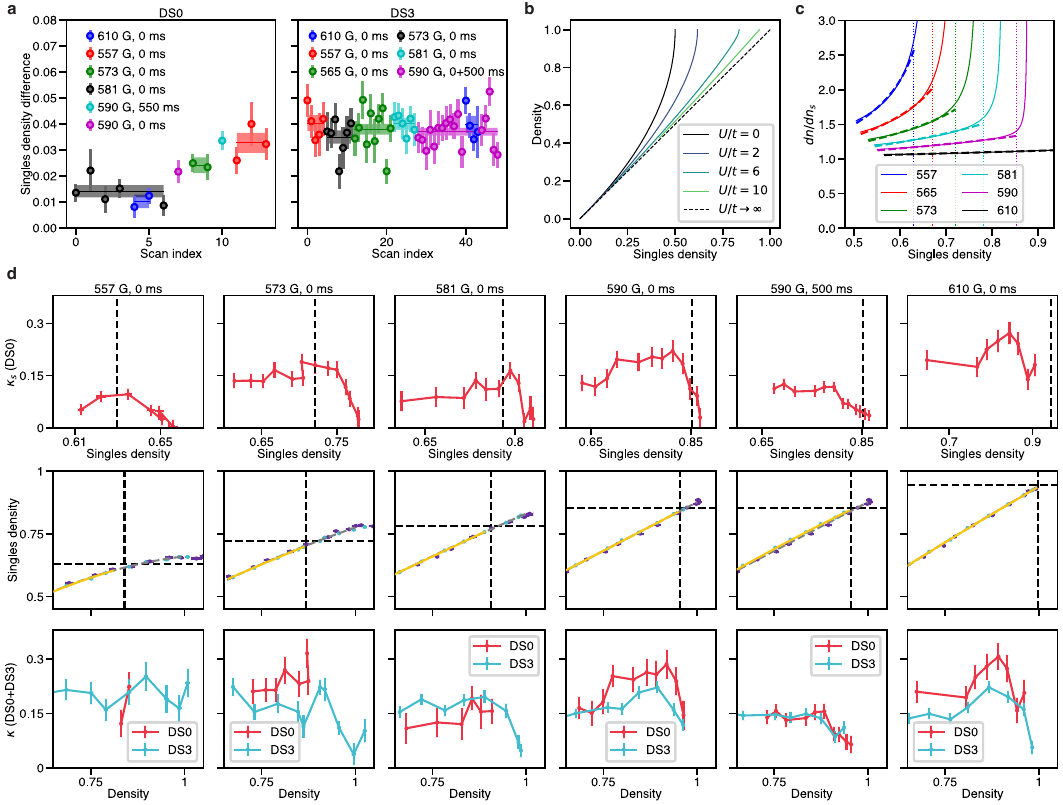"}
    \caption{\textbf{Compressibility data analysis} (\textbf{a}) Singles density difference imprinted by the step $\Delta\mu$ in calibration data at $U/t=\Ucal$ in datasets DS0 and DS3.
    We observe significant drifts in this quantity in DS0, but not in DS3, motivating a `moving' definition of $\Delta\mu$ in DS0 but not DS3 (Section~\ref{subsec:compressibility}).
    Colors indicate different subsections of the dataset during which specific values of $U/t$ and $T/t$ were studied.
    (\textbf{b}) Density vs singles density for a range of interaction strengths, computed using DQMC at $T/t=0.5$.
    The divergence in the derivative $dn/dn_s$ near half-filling makes it challenging to use data from DS0 close to half-filling, motivating the analysis in Section~\ref{subsubsec:combination}.
    (\textbf{c}) Solid lines show the derivative $dn/dn_s$ computed using DQMC at $T/t=0.5$ at experimental interaction strengths.
    Dotted vertical lines show the `cutoff' values $n_s^{\rm{cut}}$ described in Section~\ref{subsubsec:combination}, above which we discard data from DS0.
    Dashed lines show the restricted fit described in Section~\ref{subsubsec:combination}, the accuracy of which is used to decide $n_s^{\rm{cut}}$.
    (\textbf{d}) Analysis used to combine datasets DS0 and DS3 (see Section~\ref{subsubsec:combination}). Each column shows the analysis for a different combination of interaction strength and temperature.
    Top row: $\kappa_s$ vs $n_s$ data from DS0. Black dashed line shows $n_s^{\rm{cut}}$.
    Second row: $n_s$ vs $n$ from DS3 (blue) and DSConv (purple).
    Black dashed lines show $n_s^{\rm{cut}}$ and $n^{\rm{cut}}$.
    Gold line is the quadratic fit used in the compressibility analysis that relies on $dn/dn_s$ (see Section~\ref{subsubsec:combination}).
    Gray dashed line is a quadratic fit to the full doping range, which is used in less sensitive analyses that need only $n(n_s)$ rather than its derivatives (Section~\ref{subsec:conversion}).
    Third row: inferred $dn/dn_s$ values at the $n_s$ values in DS0 satsifying $n_s<n_s^{\rm{cut}}$.
    Fourth row: inferred $\kappa$ from DS0 (red) with directly measured $\kappa$ from DS3 (blue).
    \label{fig:combination_fig}}
\end{dfigure*}

\begin{dfigure*}{SI-B1-fit}
\centering
    \noindent
    \includegraphics[width=7.2in]{"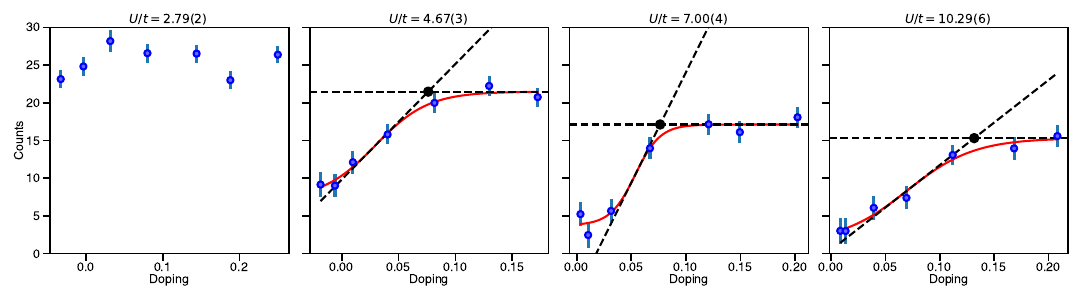"}
    \caption{\textbf{Fitting the $B_1$ pseudogap onset.} Blue markers show the low-frequency $B_1$ response vs doping and interaction, the same data as in Fig.~\ref{fig:raman_vs_U}.
    The red line is a sigmoid function fit to the data (see Section~\ref{subsec:raman_fit}).
    The black dashed lines are tangents to the fit at the sigmoid midpoint and at infinity.
    The black points show the quantity $\delta_{\star}$ defined in Section~\ref{subsec:raman_fit}.
    \label{fig:B1_fit}}
\end{dfigure*}

\begin{dfigure*}{SI-widom-fit}
\centering
    \noindent
    \includegraphics[width=6.94in]{"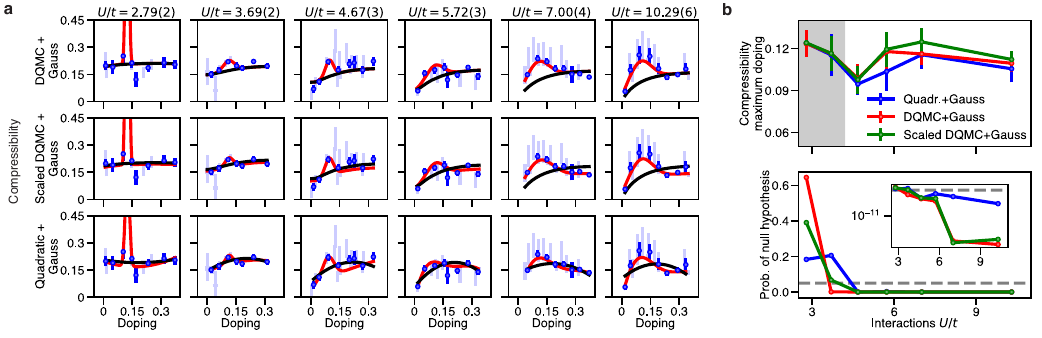"}
    \caption{\textbf{Fitting the compressibility maximum.} (\textbf{a})
    Fits to the three models described in Section~\ref{subsubsec:widom_fit}. Each fit is shown in a different row.
    Dark blue markers reproduce the data shown in Fig.~\ref{fig:widom}c.
    Pale blue points are the data shown in Fig.~\ref{fig:combination_fig}, which are binned to produce the dark blue points.
    In each row, the black curve is the null hypothesis in the likelihood ratio test described in Section~\ref{subsubsec:widom_fit}, while the red curve is the unrestricted model fit.
    The maxima of the red curves are used to extract the doping of the compressibility maximum.
    (\textbf{b}) Top panel shows the fitted doping of the compressibility maximum versus interactions using the three models. The gray dashed region indicates where the null hypothesis $p$-value for the model~\eqref{eq:widom_fit} exceeds $0.05$ (i.e. the region with no detectable maximum).
    Lower panel shows the probability of the null hypothesis used in the likelihood ratio test.
    The gray dashed line is at $0.05$.
    The inset shows the same data on a semilog plot.
\label{fig:widom_fit}}
\end{dfigure*}

\begin{dfigure*}{SI-linearity}
    \noindent
    \includegraphics[width=5.42in]{"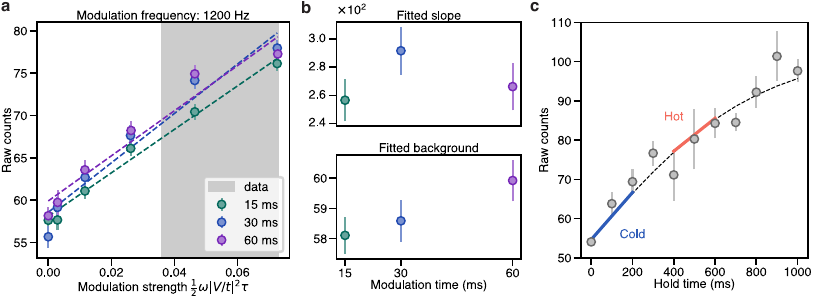"}
    \caption{\textbf{Linearity of lattice modulation response.} (a) The number of defects measured in the BI after lattice modulation of a sample near half filling at a frequency near the two magnon resonance. Dashed lines are linear fits, and the gray shaded region indicates the modulation strengths at which the data in this work were obtained. (b) The fitted slope of the modulation response indicates the expected scaling of heating with modulation amplitude $V$ and time $t$. We observe a small, approximately linear, increase in background counts associated with the duration of the modulation procedure.
    (c) Number of defects measured in the BI after holding in the physics lattice depth.
    The slope of counts versus hold time decreases with increasing heating both because the number of defects is a saturating function of the BI entropy, and because the temperature of the modulated state (which sets the conversion from absorbed energy to entropy) is higher.
    However, for the amount of modulation applied in the measurements, we remain in an approximately linear regime.
    For the high temperature measurement after 500ms hold time shown in Fig~\ref{fig:u8B1}(c), the counts after modulation are normalized by the slope at 500ms.
    \label{fig:linearity}}
\end{dfigure*}

\begin{dfigure*}{SI-driven-corr}
    \noindent
    \includegraphics[width=4.45in]{"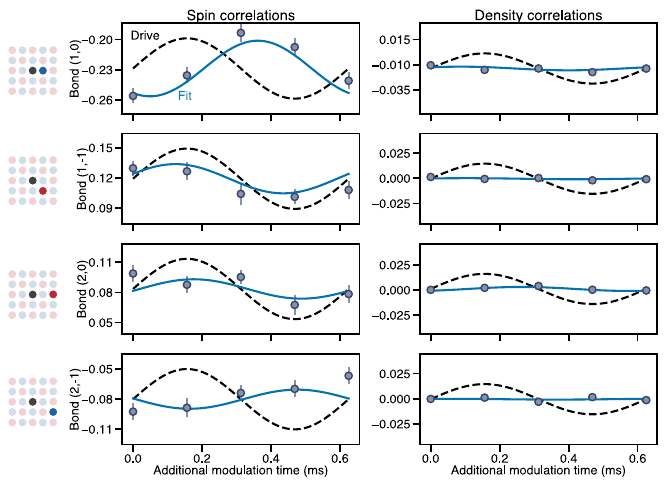"}
    \caption{\textbf{Driven correlations.} Spin and density correlations of different bonds as a function of additional modulation time after an initial $3.75~$ms of modulation, measured near half filling. Dashed lines indicate the phase of the applied drive, solid lines are sinusoidal fits at the applied drive frequency of $\omega = 2\pi\times1.6~$kHz, near the two magnon resonance.
    \label{fig:driven-corr}}
\end{dfigure*}

\begin{dfigure*}{SI-u8B1}
    \noindent
    \includegraphics[width=5.95in]{"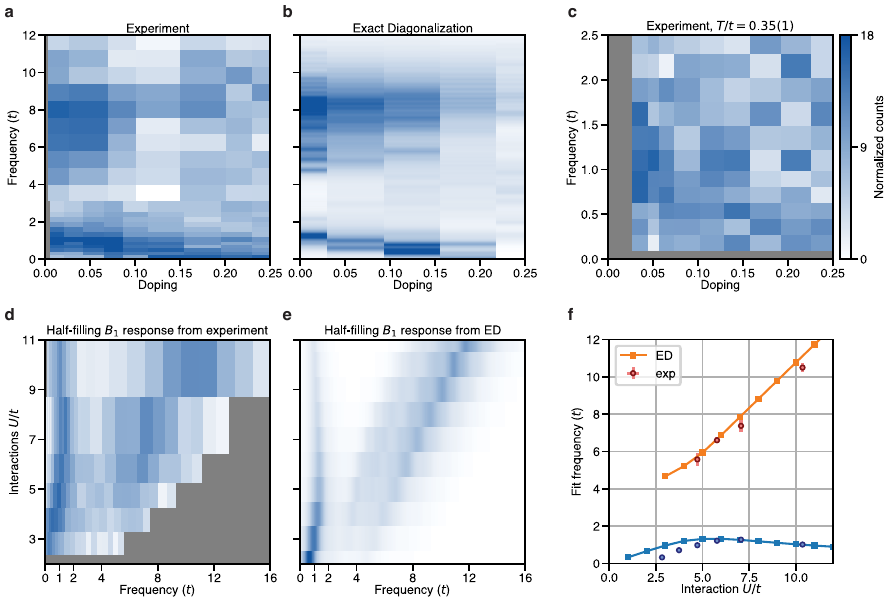"}
    \caption{\textbf{B1 response} (a) Experimentally measured $B_1$ response at $U/t=\Uviii$ at frequencies up to $12 t$. A broad $U-$scale response is seen at half-filling, which extends out to a doping of $\sim 0.20$. 
    (b) Numerically computed $B_1$ response in the ground state on a $4\times4$ finite cluster at $U/t=7$ using Exact Diagonalization (ED). The response qualitatively agrees with the experiment, although the doping resolution is limited in the numerics do to the small system size.
    (c) Experimentally measured $B_1$ response at $U/t=\Uviii$ and high temperature of $T/t=0.35(1)$ for frequencies below $2.5t$. Overall the response is much weaker than at low temperatures. Additionally, the two-magnon feature seen prominently at low temperatures is almost absent at high temperatures, and the pseudogap at low energies is less pronounced, if present at all.
    (d) Half-filling $B_1$ response from the experiment for various interaction strengths. The two-magnon peak shows non-monotonic behavior with $U$ while the $U-$scale peak is approximately linear with $U$.
    (e) same as (d) but computed numerically.
    (f) Fitted frequency of the half-filling two-magnon peaks and $U-$scale peaks from the experiment and from ED. The numerical method overestimates the two-magnon frequency at low $U$ as compared to the experiment. 
    \label{fig:u8B1}}
\end{dfigure*}

\begin{dfigure}{SI-bandgap}
    \noindent
    \includegraphics[width=3in]{"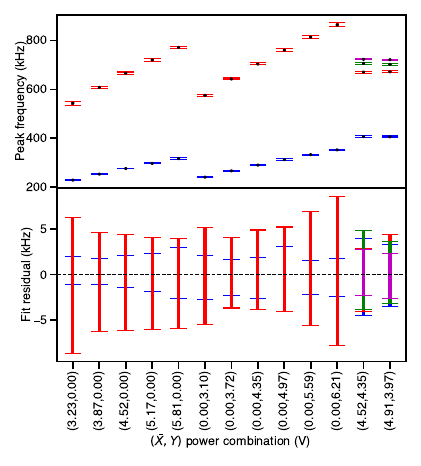"}
    \caption{\textbf{Bandgap spectroscopy.} 
    Spectroscopic data used to calibrate the lattice. The upper panel shows peak frequencies measured under a variety of lattice power conditions (labeled on the horizontal axis, as measured in Volts on a photodiode).
    Different points at the same horizontal position correspond to different resonances at the same condition, e.g. in-plane vs out-of-plane excitations.
    The lower panel shows the residuals from the fit described in Section~\ref{subsec:calibration}.\label{fig:bandgap}}
\end{dfigure}

\begin{dfigure}{SI-T-vs-U}
    \noindent
    \includegraphics[width=3in]{"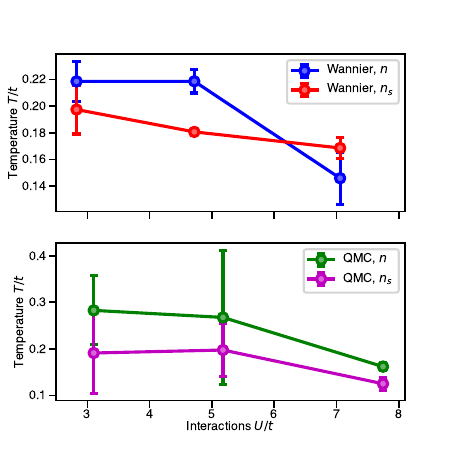"}
    \caption{\textbf{Temperature versus interactions.} Temperature versus interaction strengths, inferred from the four methods described in Section~\ref{subsec:calibration}. We use $(U/t)_{\rm{Wannier}}$ as the $x$ axis in the upper panel, and $(U/t)_{\rm{QMC}}$ in the lower.\label{fig:T_vs_U}}
\end{dfigure}

\end{document}